\documentclass[reqno,11pt]{amsart}
\usepackage{graphicx}
\usepackage{amscd}
\usepackage{slashed}
\usepackage{amssymb}
\usepackage[mathscr]{eucal}
\numberwithin{equation}{section}
\allowdisplaybreaks[1]

\newcommand{\SetFigFont}[3]{}

\title[On the Initial Value Problem for Causal Variational Principles]{On the Initial Value Problem for \\
Causal Variational Principles}

\author[F.\ Finster]{Felix Finster}
\address{Fakult\"at f\"ur Mathematik \\ Universit\"at Regensburg \\ D-93040 Regensburg \\ Germany}
\email{finster@ur.de}
\author[A.\ Grotz]{Andreas Grotz \\ \\ March 2013}
\address{Department of Mathematics \\ Harvard University \\ Cambridge, MA 02138 \\ USA}
\email{agrotz@math.harvard.edu}

\newtheorem{Def}{Definition}[section]
\newtheorem{Thm}[Def]{Theorem}
\newtheorem{Prp}[Def]{Proposition}
\newtheorem{Lemma}[Def]{Lemma}
\newtheorem{Corollary}[Def]{Corollary}

\newtheorem{Example}[Def]{Example}

\newcommand{\Thanks}{\vspace*{.5em} \noindent \thanks}
\newcommand{\beq}{\begin{equation}}
\newcommand{\eeq}{\end{equation}}
\newcommand{\Proof}{\begin{proof}}
\newcommand{\QED}{\end{proof} \noindent}
\newcommand{\QEDrem}{\ \hfill $\Diamond$}

\newcommand{\la}{\langle}
\newcommand{\ra}{\rangle}

\newcommand{\R}{\mathbb{R}}
\newcommand{\1}{\mbox{\rm 1 \hspace{-1.05 em} 1}}

\newcommand{\N}{\mathbb{N}}

\renewcommand{\H}{\mathscr{H}}
\newcommand{\U}{{\rm{U}}}

\newcommand{\bep}{\begin{pmatrix}}
\newcommand{\enp}{\end{pmatrix}}
\newcommand{\bca}{\begin{cases}}
\newcommand{\eca}{\end{cases}}

\newcommand{\F}{{\mathscr{F}}}
\newcommand{\D}{{\mathscr{D}}}

\newcommand{\M}{{\mathbb{M}}}
\newcommand{\B}{{\mathbb{B}}}
\newcommand{\Bm}{{\mathscr{B}}}

\renewcommand{\L}{{\mathcal{L}}}
\newcommand{\Sact}{{\mathcal{S}}}
\newcommand{\s}{{\mathfrak{s}}}

\newcommand{\Lin}{\text{\rm{L}}}
\newcommand{\I}{{\mathfrak{I}}}
\newcommand{\J}{{\mathscr{J}}}
\newcommand{\K}{{\mathscr{K}}}

\DeclareMathOperator{\supp}{supp}

\newcommand{\bpm}{\begin{pmatrix}}
\newcommand{\epm}{\end{pmatrix}}

\begin{document}

\begin{abstract}
We formulate the initial value problem for causal variational principles
in the continuous setting on a compact metric space.
The existence and uniqueness of solutions is analyzed.
The results are illustrated by simple examples.
\end{abstract}

\maketitle

\tableofcontents

\section{Introduction}
Causal variational principles arise in the context of relativistic quantum theory
(see the survey article~\cite{rrev} and the references therein).
In~\cite{continuum} they were introduced from a mathematical perspective,
and the existence of minimizers has been proven in various situations.
A more detailed analysis of causal variational principles and of the corresponding
Euler-Lagrange equations is carried out in~\cite{support, lagrange}.

In the present paper, we analyze the question how an initial value problem can be posed
for causal variational principles, and whether it has a unique solution.
For technical simplicity, we restrict attention to the so-called {\em{continuous}} setting on a
{\em{compact}} manifold as introduced in~\cite[Section~1.4]{continuum}
and more generally in~\cite{support}.
But using the techniques in~\cite{lagrange}, many methods and results could
be extended in a straightforward way to the non-compact setting.
Since we shall not make use of the manifold structure, we now let~$\F$ be a
compact {\em{metric space}}\footnote{We remark for clarity that all our results
hold just as well for a compact metrizable topological space (i.e., in view of the
Urysohn metrization theorem, for a second-countable compact Hausdorff space).
In fact, we never make use of the metric, but work exclusively with the topology.
When referring to results on metric spaces, one can simply work with an
arbitrarily chosen metric.}.
For a given {\em{Lagrangian}}~$\L \in C^0(\F \times \F, \R^+_0)$ which is symmetric
(i.e.\ $\L(x,y)=\L(y,x)$ for all~$x,y \in \F$),
we introduce the {\em{action}}~$\Sact$ by
\beq \label{Sdef}
\Sact[\rho] = \iint_{\F \times \F} \L(x,y)\: d\rho(x)\: d\rho(y)\:.
\eeq
Here~$\rho$ is a normalized positive regular Borel measure on~$\F$, referred to as the
{\em{universal measure}}. Our action principle is to minimize~$\Sact$ by varying~$\rho$ in the class
\beq \label{rhoclass}
\M^+_1(\F) := \{ \text{normalized positive
regular Borel measures on~$\F$} \}\:.
\eeq
For a minimizer~$\rho$, space-time is defined as the support of~$\rho$,
\[ M := \supp \rho \subset \F \:. \]
A-priori, $M$ is a topological space (carrying the relative topology of~$\F$).
Additional structures, like the following causal structure, are induced on~$M$ by the Lagrangian.
\begin{Def}  Two space-time points~$x, y \in M$ are called
{\textbf{time-like}} separated if~$\L(x,y)>0$. 
They are called {\textbf{space-like}} separated if~$\L(x,y)=0$.
\end{Def} \noindent
For more space-time structures in the richer context of causal fermion systems
we refer to~\cite{rrev, lqg}.

For the following concepts, it is important to keep in mind that space-time is not a-priori given, but emerges
by minimizing the causal action. When varying~$\rho$, one also varies space-time together with
all the additional space-time structures.
This situation can be understood similar as in general relativity, where the space-time manifold
with its Lorentzian metric and causal structure are not a-priori given, but are obtained dynamically
by solving the Einstein equations.

When solving the classical Cauchy problem,
instead of searching for a global solution, it is often easier to look for a local solution
around a given initial value surface. This concept of a local solution also reflects the
common physical situation where the physical system under consideration is only a small
subsystem of the whole universe.
With this in mind, our first step is to ``localize'' our variational principle by introducing
the so-called {\em{inner variational principle}}. 
To this end, we fix a Borel subset~$\I \subset \F$ (the ``inner system'')
and minimize the action
\beq \label{SIdef}
\Sact_\I[\rho, \phi] := \iint_{\I \times \I} \L(x,y)\: d\rho(x)\: d\rho(y)
+ 2 \int_{\I} (\phi(x) - \s) \: d\rho(x)
\eeq
under variations in the class
\beq \label{rhoIclass}
\M^+(\I) := \{ \text{bounded positive regular Borel measures on~$\I$} \}\:,
\eeq
where~$\s>0$ is a parameter, and~$\phi$ is a non-negative function in the class
\beq \label{phiclass}
\B^+(\I) := \big\{ \phi:\I\rightarrow \R^+_0 \,\big|\, \text{$\phi$ bounded and lower semi-continuous} \big\} \:.
\eeq
The derivation of the inner variational principle will be given in Section~\ref{sec-inner}.
Here we only explain the basic concepts behind the inner variational principle.
First, it is important to observe that the causal variational principle~\eqref{Sdef}
is {\em{non-local}} in the sense that~$\L(x,y)$ may be non-zero even for points~$x,y$ which are
are far apart. This means that the subsystem~$\I$ will be influenced also by the
universal measure outside this subsystem.
This effect is taken into account in~\eqref{SIdef} by the function~$\phi$,
referred to as the {\em{external potential}}.
The parameter~$\s$, on the other hand, plays the role of a Lagrange multiplier
that takes care of the volume constraint in~\eqref{rhoclass} 
(note that the measure in~\eqref{rhoclass} is normalized, whereas the measure in~\eqref{rhoIclass}
is not).

The external potential is closely linked to our concept for prescribing initial values, as we now explain.
In the setting of causal variational principles, initial values are introduced naturally
by prescribing a measure~$\rho_0 \in \M^+(\I)$ (the ``initial data'')
and by demanding that~$\rho \geq \rho_0$.
If we implemented the inequality~$\rho \geq \rho_0$ as a side condition for the
inner variational principle, treating the inequality constraint with Lagrange multipliers
would give rise to additional terms in the EL equations.
This means that the EL equations would depend on the initial data,
in clear contrast to the usual concept of solving a-priori given EL equations for prescribed
initial data. For this reason, imposing the side condition~$\rho \geq \rho_0$ is not a
sensible concept. It is preferable to minimize~\eqref{SIdef} without
constraints, but to choose~$\phi$ in such a way that the minimizing measure~$\rho$
satisfies the inequality~$\rho \geq \rho_0$. This leads to the following definition.
\begin{Def} \label{defcauchy}
Given a measure~$\rho_0 \in \M^+(\I)$ and a parameter~$\s>0$,
a measure~$\rho \in \M^+(\I)$ is called a
{\bf{solution of the initial value problem}} in~$\I$ with initial data~$\rho_0$
and external potential~$\phi \in \B^+(\I)$
if it is a minimizer of the inner variational principle with the additional property~$\rho \geq \rho_0$.
We denote the set of solutions together with the corresponding external potentials by
\beq \label{solset}
\mathfrak{S}_\I(\rho_0) = \big\{ (\rho, \phi) \,\big|\,\text{$\rho$ solves the initial value problem
with external potential~$\phi$} \big\}\,.
\eeq
\end{Def} \noindent
A detailed discussion of our method for prescribing initial data will be given in Section~\ref{sec-initdata}.

Note that in the above definition, the external potential can be chosen arbitrarily up to the requirement that the corresponding solution of the inner variational principle should comply with the initial data.
Following the concept that the external potential describes the influence of the outer system,
choosing~$\phi$ can be viewed as suitably ``preparing'' the outer system in such a way
that the resulting universal measure is compatible with the initial data.
Since we cannot expect that there is a unique way of preparing the outer system,
there may be a whole family of possible choices of~$\phi$.
As the outer system is unknown for principal reasons, the choice of~$\phi$ is
not determined a-priori, and it is not unique.
As an example, for a given minimizer~$(\phi, \rho)$, increasing the external potential outside
the support of~$\rho$ does not change the action~\eqref{SIdef} and clearly preserves the
minimizing property of a measure~$\rho$.
As an additional difficulty, even for a fixed external potential, in general there will be no
unique minimizing measure~$\rho$.
Despite these complications, we succeed in constructing a uniquely defined so-called {\em{domain of dependence}}
on which the minimizing measure~$\rho$ is unique for any choice of~$\phi$.
Moreover, we construct a so-called {\em{maximal optimal solution}}
where both~$\rho$ and~$\phi|_{\supp \rho}$ are uniquely determined
by suitably ``optimizing'' the external potential.

The paper is organized as follows.
In Section~\ref{sec-setup}, we derive the inner variational principle as well as the corresponding Euler-Lagrange equations. Moreover, we discuss 
our method of prescribing initial data and introduce different notions of optimal solutions. 
In Section~\ref{sec-existence}, we prove existence results for the inner variational principle. 
Moreover, we characterize those initial data which admit solutions of the initial value problem, and we prove existence of optimal solutions. In Section~\ref{sec-uniqueness}, we introduce the domain of dependence
as the largest set where the inner variational principle has a unique solution for
every~$\phi \in \B^+(\I)$. Furthermore, we analyze the uniqueness of optimal solutions
and construct the uniquely determined maximal optimal solution.
Finally, Section~\ref{sec-examples} provides some simple yet instructive examples of initial value problems.

\section{Setting up the Initial Value Problem} \label{sec-setup}

\subsection{The Inner Variational Principle}\label{sec-inner}
The universal measure~$\rho$ in~\eqref{Sdef} should be regarded as describing the whole space-time.
In most applications, however, one is interested only in a subregion of space-time
whose volume is much smaller than the total volume of space-time.
In order to describe this situation, we now ``localize'' the variational principle~\eqref{Sdef} as follows.
By rescaling the measure~$\rho$ we arrange that~$\rho(\F)=V$ with~$V>0$
(this is useful because we will later take the infinite volume limit~$V \rightarrow \infty$).
Moreover, we fix a Borel subset~$\I \subset \F$ (the {\em{inner system}})
and decompose the measure~$\rho$ as
\[ \rho = \rho_\text{in} + \rho_\text{out} \]
with~$\rho_\text{in} = \chi_\I \,\rho$ and~$\rho_\text{out} = \chi_{\F \setminus \I}\, \rho$
(and~$\chi$ denotes the characteristic function).
We also set~$V_\text{in} = \rho_\text{in}(\F) = \rho(\I)$ and~$V_\text{out} = \rho_\text{out}(\F)$;
clearly~$V=V_\text{in} + V_\text{out}$. The action~\eqref{Sdef} becomes
\beq \label{Sdecomp}
\Sact[\rho] = \Sact[\rho_\text{in}] + \Sact[\rho_\text{out}] + 2
\iint_{\F \times \F} \L(x,y)\: d\rho_\text{in}(x)\: d\rho_\text{out}(y) \:.
\eeq
We have the situation in mind that only~$\rho_\text{in}$ is known, whereas the measure~$\rho_\text{out}$
in the ``outer system''~$\F \setminus \I$ is inaccessible to the physical system under consideration.
This means that, in order to derive the effective action principle of the inner system, we
only consider variations of the measure~$\rho_\text{in}$. 
It is important to notice that the volume~$\rho_\text{in}(\F)$ need not be preserved in the
variation, as only the total volume of the whole space-time must be kept fixed.
The latter can be arranged by rescaling~$\rho_\text{out}$. Thus for a variation~$\tilde{\rho}_\text{in}$
of~$\rho_\text{in}$ we consider the corresponding variation of~$\rho$ as given by
\[ \tilde{\rho} = \tilde{\rho}_\text{in} + \frac{V - \tilde{V}_\text{in}}{V_\text{out}}\: \rho_\text{out} \:, \]
where we impose that~$\supp \tilde{\rho}_\text{in} \subset \I$
and set~$\tilde{V}_\text{in} := \tilde{\rho}_\text{in}(\F)$.
The corresponding action~\eqref{Sdecomp} becomes
\[ \Sact[\tilde{\rho}] = \Sact[\tilde{\rho}_\text{in}] + \Big( \frac{V - \tilde{V}_\text{in}}{V_\text{out}} \Big)^2\:
\Sact[\rho_\text{out}] + 2\:\frac{V - \tilde{V}_\text{in}}{V_\text{out}}
\iint_{\F \times \F} \L(x,y)\: d\tilde{\rho}_\text{in}(x)\: d\rho_\text{out}(y) \:. \]
We now consider the limiting case when the total volume~$V \rightarrow \infty$, whereas~$V_\text{in}$
and~$\tilde{V}_\text{in}$ stay bounded\footnote{In order to make this limit formally rigorous,
one could consider a sequence~$(\F_n)$ of metric spaces together with
embeddings~$\iota_n : \I \hookrightarrow \F_n$
and a sequence of suitable measures~$\rho_{\text{out}, n}$ on~$\F_n \setminus \iota_n(\I)$.}.
Moreover, we assume that~$\Sact[\rho_\text{out}]$ grows linearly in~$V_\text{out}$, so that the following
limits exist,
\begin{align*}
\phi(x) &:= \lim_{V_\text{out} \rightarrow \infty} \int_\F \L(x,y)\: d\rho_\text{out}(y) \in C^0(\I, \R^+_0) \\
\s &:= \lim_{V_\text{out} \rightarrow \infty} \frac{\Sact[\rho_\text{out}]}{V_\text{out}}\:.
\end{align*}
Under these assumptions, our action converges after subtracting an irrelevant constant,
\[ \lim_{V \rightarrow \infty} \Big( \Sact[\tilde{\rho}] - \Sact[\rho_\text{out}] \Big)
= \Sact[\tilde{\rho}_\text{in}] + 2 \int_{\I} \phi(x)\: d\tilde{\rho}_\text{in} - 2\, \tilde{V}_\text{in}\: \s \:. \]
To simplify the notation, we again denote the measure~$\tilde{\rho}_\text{in}$ by~$\rho$.
We then obtain the action~\eqref{SIdef}, 
which is to be minimized under variations in the class~\eqref{rhoIclass}.
This variational principle can be regarded as a generalization of our original action principle~\eqref{Sdef} and~\eqref{rhoclass}, where we replaced the normalization constraint in~\eqref{rhoclass} by the Lagrange multiplier term~$-2 \s \rho(\I)$.
As indicated in the introduction, the influence of the universal measure in the outer system is described effectively in the inner action~\eqref{SIdef} by the external potential~$\phi$.
In view of our later constructions, it is useful to allow the external potential to be in the larger class $\B^+(\I)$ of lower semi-continuous functions (see~\eqref{phiclass}).

Obviously, in the case~$\s <0$ the variational principle only has the trivial minimizer~$\rho=0$.
In the case~$\s = 0$, every minimizing measure is supported on the zero set of~$\phi$, and
restricting attention to measures with this property, the action~\eqref{SIdef} reduces to~\eqref{Sdef}.
In order to rule out these trivial cases, we shall always assume that~$\s > 0$.
We thus obtain the following variational principle.
\begin{Def} Let~$\I$ be a metric space and~$\L \in C^0(\I \times \I, \R^+_0)$ a symmetric Lagrangian.
Given a parameter~$\s>0$ and a function~$\phi \in \B^+(\I)$, the {\bf{inner variational principle}}
is to minimize the functional~$\Sact_\I$ in \eqref{SIdef} under variations of~$\rho$ in the class~$\M^+(\I)$. 
\end{Def}
We will see in Section~\ref{sec-existinner} that the inner variational principle has solutions for any~$\s>0$ and~$\phi\in\B^+(\I)$.

\subsection{The Euler-Lagrange Equations} \label{sec-euler}
In this section, we derive the Euler-Lagrange (EL) equations corresponding to the
inner variational principle. For a convenient notation, we set
\[ \M(\I) := \{ \text{bounded signed regular Borel measures on $\I$} \} \]
and introduce the short notations
\begin{align*}
\L \mu (x) &= \int_\I \L(x,y)\: d\mu(y) \\
\la f, \mu \ra &= \int_\I f(x)\: d\mu(x) \:, 
\end{align*}
where~$\mu\in\M(\I)$ is any signed measure and~$f:\I\rightarrow\R$ a measurable function.
Then the action $\Sact_\I$ can be written in the compact form
\beq \Sact_\I[\rho, \phi] = \la \L \rho, \rho \ra + 2\, \la \phi-\s, \rho \ra \:. \label{act-shorthand} \eeq
In order to further simplify the setting, we note that the action~\eqref{SIdef} is invariant under
the rescaling
\beq \label{scale}
\phi \rightarrow \lambda\, \phi \:,\qquad \rho \rightarrow \lambda\, \rho \:,\qquad
\s \rightarrow \lambda\, \s \:,\qquad
\Sact_\I \rightarrow \lambda^{-2}\: \Sact_\I  \qquad (\text{where~$\lambda > 0$})\:.
\eeq
With this rescaling, we can always arrange that~$\s=1$.

In the next proposition we use similar methods as in~\cite[Section~3.1]{support}.
\begin{Prp} {\bf{(Euler-Lagrange Equations)}} \label{prp-euler}
Every minimizer~$\rho$ of the inner variational principle has the properties
\begin{align}
&\L \rho + \phi \big|_{\supp \rho} \equiv \min_\I \big( \L \rho + \phi \big) = 1 \quad \text{and} \label{euler1} \\
&\la \L \mu , \mu \ra \geq 0 \qquad
\text{for all~$\mu \in \M(\supp \rho)$}\:. \label{euler2}
\end{align}
\end{Prp} \noindent
\Proof
Let~$\rho\in\M^+(\I)$ be a minimizer of the inner variational principle~\eqref{SIdef}
with external potential~$\phi\in\B^+(\I)$.
We consider the family of measures~$\tilde{\rho}_t = \rho + t \,\delta_x$ with~$t \geq 0$, where~$\delta_x$
denotes the Dirac measure supported at~$x \in \I$. Taking the right-sided derivative of~$\Sact_\I[\tilde{\rho}_t,\phi]$
with respect to~$t$, we find that
\beq \label{in1}
(\L \rho)(x) + \phi(x) - 1 \geq 0 \qquad \text{for all~$x \in \I$}\:.
\eeq
Next, we consider for~$t \in (-1,1)$ the family of measures~$\tilde{\rho}_t = (1+t)\, \rho$.
Again differentiating the action with respect to~$t$, we find that
\beq \label{in2}
\int_\I \big( (\L \rho)(x) + \phi(x) - 1 \big)\: d\rho(x) = 0 \:.
\eeq
Combining~\eqref{in1} and~\eqref{in2} gives~\eqref{euler1}. 

In order to prove~\eqref{euler2}, we define the real Hilbert space~$\H_\rho = L^2(\I, d\rho)$ and the linear operator
\beq \L_\rho \::\: \H_\rho \rightarrow \H_\rho\:,\qquad (\L_\rho \psi)(x) =
\int_\I \L(x,y)\: \psi(y)\: d\rho(y) \:. \label{Lrho-op} \eeq
For any bounded function~$\psi \in \H_\rho$, we consider
for~$t \in (-\varepsilon, \varepsilon)$ (with~$0 < \varepsilon < 1/\|\psi\|_\infty$) the family of measures
\[ \tilde{\rho}_t = (1 + t \,\psi)\: \rho \]
(where~$\psi \rho$ is the signed measure~$(\psi \rho)(\Omega) := \int_\Omega \psi d\rho$).
Differentiating the action twice, we obtain
\[ 0 \leq \iint_{\I \times \I} \L(x,y)\: \psi(x) \,d\rho(x)\: \psi(y) \,d\rho(y) = \la \psi, \L_\rho \psi \ra_\H\:, \]
which shows that $\L_\rho$ is a positive semi-definite operator on~$\H_\rho$.
The relation~\eqref{euler2} follows by approximating any given measure~$\mu\in\M(\supp(\rho))$
by a series~$(\mu_n)$ of measures of the form~$\mu_n = \psi_n \,\rho$ with~$\psi_n \in \H_\rho$.
\QED
Inserting the relation~\eqref{euler1} into the action~\eqref{act-shorthand}, we obtain the following result.
\begin{Corollary} \label{cor-minact}
For a minimizer~$\rho\in\M^+(\I)$ of the inner variational principle with external potential~$\phi\in\B^+(\I)$, the inner action takes the value
\beq \Sact[\rho, \phi] = - \la \L \rho, \rho \ra = \la \phi-1, \rho \ra \,. \label{minact} \eeq
\end{Corollary}

\subsection{Prescribing Initial Data} \label{sec-initdata}
To motivate our method, let us assume that we want to find a minimizer~$\rho$ of
the inner variational principle which has the additional property that~$\rho \geq \rho_0$
for a given measure~$\rho_0 \in \M^+(\I)$ (the ``initial data'').
The most obvious idea for implementing 
the constraint~$\rho \geq \rho_0$ is to write~$\rho$ in the form~$\rho = \rho_0 + \nu$
with a measure~$\nu \in \M^+(\I)$. Substituting this ansatz into~\eqref{SIdef}, one obtains
\[ \Sact = \iint_{\I \times \I} \L(x,y)\: d\nu(x)\: d\nu(y)
+ 2 \int_{\I} (\tilde{\phi}(x) - 1) \: d\nu(x) + \text{const} \:, \]
where the new external potential~$\tilde{\phi}$ is given by
\[ \tilde{\phi}(x) = \phi(x) + \int_\I \L(x,y)\: d\rho_0 \:. \]
Thus one can minimize~$\Sact$ under variations of~$\nu \in \M^+(\I)$.
According to Proposition~\ref{prp-euler}, we obtain the EL equations
\begin{align}
&\L \nu + \tilde{\phi} \big|_{\supp \nu} \equiv \min_\I \big( \L \nu + \tilde{\phi} \big) = 1 \quad \text{and} \label{nu1} \\
&\la \L \mu , \mu \ra \geq 0 \qquad
\text{for all~$\mu \in \M(\supp \nu)$}\:. \label{nu2}
\end{align}
The problem is that the EL equations~\eqref{nu1} and~\eqref{nu2} are considerably
weaker than the earlier equations for~$\rho$, \eqref{euler1} and~\eqref{euler2},
because they must hold only on~$\supp \nu$,
but not on~$\supp \rho$ (note that in general~$\supp \rho \supsetneq \supp \nu$).
For this reason, minimizing~$\nu$ is {\em{not}} the correct procedure.
Instead, our strategy is to minimize~$\rho$ over the whole class~$\M^+(\I)$, but to always
choose the external potential in such a way that the minimizer satisfies the constraint~$\rho \geq \rho_0$.
This leads to the initial value problem formulated in Definition~\ref{defcauchy} above.

For the applications, it might be useful to consider more general initial data, which consists of the measure~$\rho_0$
and in addition of a closed subset~$\I_0 \subset \I$. We demand that the conditions~\eqref{euler1}
and~\eqref{euler2} also hold
on the set~$\I_0$.
\begin{Def} \label{defcauchy2}
Given a measure~$\rho_0 \in \M^+(\I)$ and a closed set~$\I_0 \subset \I$,
a measure~$\rho \in \M^+(\I)$ is called a
{\bf{solution of the initial value problem}} in~$\I$ with initial data~$(\rho_0, \I_0)$
and external potential~$\phi \in \B^+(\I)$
if it is a minimizer of the inner variational principle with the following additional properties:
\begin{itemize}
\item[(a)] $\rho \geq \rho_0$ \\[-1em]
\item[(b)] $\L \rho + \phi \big|_{\I_0} \equiv 1$
\item[(c)] 
$ \displaystyle \la \L \mu, \mu \ra \geq 0 \qquad
\text{for all~$\mu \in \M(\I_0 \cup \supp \rho)$}$\:.
\end{itemize}
In analogy to~\eqref{solset}, we denote the set of solutions by $\mathfrak{S}_\I(\rho_0,\I_0)$.
\end{Def} 
The initial value problem in Definitions~\ref{defcauchy} and~\ref{defcauchy2} cannot be solved for arbitrarily chosen initial data~$(\rho_0,\I_0)$. For example, if the measure~$\rho_0$ is chosen such that there is a
point~$x\in\supp(\rho_0)$ with~$(\L\rho_0)(x)>1$, then the EL equation~\eqref{euler1} excludes
existence of a minimizer~$\rho\in\M^+(\I)$ with~$\rho\geq\rho_0$ for any external potential.
In Section~\ref{sec-existinit} we will characterize those initial data which admit solutions of the initial
value problem.

\subsection{Optimizing the External Potential} \label{sec-optpot}
Let us assume that the initial data~$\rho_0$ or~$(\rho_0,\I_0)$ admits a solution~$(\rho, \phi)$ of the initial value problem.
Then this solution will in general not be unique. 
Moreover, there is an arbitrariness in choosing the external potential.
Our strategy for getting uniqueness is to choose the external potential in an ``optimal way''. There are
three basic notions of optimality:
\beq \label{optimize1}
\begin{array}{cl}
\text{(A)} & \text{Minimize the action~$\Sact_\I[\rho, \phi]$, where~$(\rho,\phi) \in \mathfrak{S}_\I(\rho_0,\I_0)$.} \\[0.2em]
\text{(B)} & \text{Minimize the value of~$\max_{\supp(\rho)} \phi$, where~$(\rho,\phi) \in \mathfrak{S}_\I(\rho_0,\I_0)$.} \\[0.2em]
\text{(C)} & \text{Maximize the volume~$\rho(\I)$, where~$(\rho,\phi) \in \mathfrak{S}_\I(\rho_0,\I_0)$.}
\end{array}
\eeq
For clarity, we note that whether to maximize or to minimize in the above optimization problems
is determined by the requirement to avoid trivial minimizers.
Namely, if we had taken the reverse choice in any of the problems~(A)--(C), 
one verifies immediately from~\eqref{euler1} and~\eqref{minact}
as well as from the condition~$\rho\geq\rho_0$ that the measure~$\rho=\rho_0$ would be a trivial solution.

Solutions exist for all of the optimization problems in~\eqref{optimize1} (see Theorem~\ref{thm-exopt}
in Section~\ref{sec-existopt}), but neither $\rho$ nor $\phi$ are unique in general (cf.\ the examples in
Section~\ref{sec-examples}).
Therefore, we propose another notion of optimality by first maximizing the volume and then maximizing the action:
\beq\label{optimize2}
\text{(D)} \quad \text{Maximize the action~$\Sact_\I[\rho, \phi]$ in $\mathfrak{S}_{\I}^\text{maxV}(\rho_0,\I_0)$}	\:,
\eeq
where $\mathfrak{S}_{\I}^\text{maxV}(\rho_0,\I_0)$ is defined as the set of solutions of the initial value problem with maximal volume,
\[ \mathfrak{S}_{\I}^\text{maxV}(\rho_0,\I_0):=
\bigg\{ (\rho,\phi)\in\mathfrak{S}_\I\,\Big|\, \rho(\I)=\max_{(\tilde{\rho},\tilde{\phi})\in\mathfrak{S}_\I} \tilde{\rho}(\I) \bigg\}\:. \]
In Section~\ref{sec-uniqueopt} we will prove that solving the optimization problem~(D) in suitable
space-time regions will indeed give a unique solution of the initial value problem.
This analysis will also explain why in~(D) we must maximize (and not minimize) the action.

\section{Existence Results} \label{sec-existence}
We now enter the analysis of the inner variational principle and of solutions of the initial value problem.
We always keep~$\rho_0 \in \M^+(\I)$ fixed and use the rescaling~\eqref{scale}
to set~$\s=1$. 
For the existence results in this section we need to assume that the inner system~$\I$ is a closed subset of~$\F$ and that the Lagrangian~$\L$ is strictly positive on the diagonal,
\beq \label{posdiag}
\L(x,x) > 0 \qquad \text{for all~$x \in \I$}\:.
\eeq
\subsection{Preparatory Considerations} \label{sec-prepconsider}
The following simple observation makes it possible to construct new minimizers
from a given minimizer of the inner variational principle.
\begin{Lemma}
Suppose that~$\rho$ is a minimizer of~$\Sact_\I$ with external potential~$\phi\in\B^+(\I)$. Then
any measure~$\tilde{\rho} \in \M^+(\I)$ with~$\tilde{\rho} \leq \rho$ is a minimizer of~$\Sact_\I$
with external potential~$\tilde{\phi}\in\B^+(\I)$ given by
\[ \tilde{\phi}(x) = \phi(x) + \int_\I \L(x,y)\: d(\rho-\tilde{\rho})(y)\:. \]
\end{Lemma}
\Proof We first note that
$$\L\tilde{\rho}(x)+\tilde{\phi}(x)-1 = (\L\rho)(x)+\phi(x)-1=0 \quad\text{for all } x\in\supp\rho\,.$$
Setting $\mu:=\rho-\tilde\rho\in\M^+(\supp\rho)$, we then find that
\begin{align*}
	\Sact_\I[\rho, \tilde{\phi}] &= \Sact_\I[\tilde{\rho}+\mu, \tilde{\phi}] = \Sact_\I[\tilde{\rho}, \tilde{\phi}] +2\la \L\tilde{\rho}, \mu \ra +\la \L\mu, \mu \ra+2\la \tilde{\phi}-1, \mu \ra \\
	&\geq \Sact_\I[\tilde{\rho}, \tilde{\phi}] +2\la \L\tilde{\rho}+\tilde{\phi}-1, \mu \ra = \Sact_\I[\tilde{\rho}, \tilde{\phi}] \,.
\end{align*}
Thus for any $\nu\in\M^+(\I)$,
\begin{align*}
	\Sact_\I[\tilde{\rho}, \tilde{\phi}] \leq \Sact_\I[\rho, \tilde{\phi}] = \Sact_\I[\rho, \phi] + 2 \la \L(\rho-\tilde{\rho}), \rho \ra \leq \Sact_\I[\nu, \phi] + 2 \la \L(\rho-\tilde{\rho}), \rho \ra 
	\leq \Sact_\I[\nu, \tilde{\phi}]\,,
\end{align*}
where we used that $\rho$ is a minimizer of $\Sact_\I[\,.\,,\phi]$ and that $\phi\leq\tilde{\phi}$. 
We conclude that~$\tilde{\rho}$ is a minimizer of $\Sact_\I[\,.\,,\tilde{\phi}]$.
\QED
The previous lemma is particularly useful for ``localizing'' a solution in a closed subset of~$\I$:
\begin{Corollary} \label{cor-localize}
Suppose that~$\rho$ is a minimizer of~$\Sact_\I$ with external potential~$\phi$.
Choosing a closed subset~$\J \subset \I$, we set
\[ \tilde{\rho} = \chi_{\J} \,\rho \qquad \text{and} \qquad
\tilde{\phi}(x) = \phi(x) + \int_{\I \setminus \J} \L(x,y)\: d\rho(y) \:. \]
Then~$\tilde{\rho}$ is a minimizer of~$\Sact_\I$ with external potential~$\tilde{\phi}$.
\end{Corollary}

The next simple estimate gives some information on the support of a minimizing measure.
\begin{Lemma} \label{nlemma-supp}
Let~$\rho$ be a minimizer of~$\Sact_\I$ with external potential~$\phi$. Then
\[ \supp \rho \subseteq \{ x \in \I \:|\: \phi(x) \leq 1 \} \:. \]
\end{Lemma}
\Proof Assume conversely that there is~$\varepsilon>0$ and a set~$U \subset \I$
with~$\rho(U)>0$ and~$\phi|_U \geq 1+\varepsilon$. Then
\begin{align*}
\Sact_\I [\chi_{\I \setminus U}\, \rho, \phi] &=
\la \chi_{\I \setminus U}\, \rho, \L \,\chi_{\I \setminus U}\, \rho \ra + 2\, \la (\phi-1_\I), \chi_{\I \setminus U}\, \rho \ra \\
&\leq \la \L \rho, \rho \ra + 2 \int_{\I \setminus U} (\phi-1)\, d\rho \\
&= \Sact_\I [\rho, \phi] -2  \int_U (\phi-1)\, d\rho 
\leq \Sact_\I [\rho, \phi] -2  \, \varepsilon\, \rho(U) < \Sact_\I [\rho, \phi]\:,
\end{align*}
in contradiction to the minimality of~$\rho$.
\QED

A similar estimate allows us to modify the external potential while preserving
the minimizing property of~$\rho$.
\begin{Lemma} \label{nlemmareplace}
Let~$\rho$ be a minimizer of~$\Sact_\I$ with external potential~$\phi$. Then~$\rho$ is also a minimizer
of~$\Sact_\I$ if the external potential is replaced by any function~$\tilde{\phi} \in \B^+(\I)$ with the properties
\[ \left\{ \begin{array}{ll} \tilde{\phi}(x) = \phi(x) & \text{if $x \in \supp(\rho)$} \\[0.2em]
\tilde{\phi}(x) \geq \min \big( \phi(x), 1 \big) & \text{if $x \not\in \supp(\rho)$}\:. \end{array} \right. \]
\end{Lemma}
\Proof Let~$U = \{x \in \I \:|\: \phi(x) < 1 \}$. 
Then for every~$\tilde{\rho} \in \M^+(\I)$,
\[ \Sact_\I[\rho, \tilde{\phi}] = \Sact_\I[\rho, \phi] \leq \Sact_\I[\chi_U \tilde{\rho}, \phi] \:, \]
where we used that~$\phi$ and~$\tilde{\phi}$ coincide on the support of~$\rho$ and that~$\rho$
is a minimizer. Next, we know by assumption that on the set~$U$, the inequality~$\phi \leq \tilde{\phi}$ holds,
and thus
\[ \Sact_\I[\chi_U \tilde{\rho}, \phi] \leq \Sact_\I[\chi_U \tilde{\rho}, \tilde{\phi}] \:. \]
Finally, we have the estimate
\begin{align*}
\Sact_\I[\chi_U \tilde{\rho}, \tilde{\phi}]
&= \big\la \chi_U \tilde{\rho}, \L \,\chi_U \tilde{\rho} \big\ra + 2\, \big\la (\tilde{\phi}-1_\I), \chi_U \tilde{\rho} \big\ra \\
&\leq \big\la \L \tilde{\rho}, \tilde{\rho} \big\ra + 2 \int_U (\tilde{\phi}-1_\I) \,d\tilde{\rho} \\
&= \Sact_\I[\tilde{\rho}, \tilde{\phi}] - 2 \int_{\I \setminus U} (\tilde{\phi}-1_\I) \,d\tilde{\rho} \:\leq\: \Sact_\I[\tilde{\rho},
\tilde{\phi}]\:,
\end{align*}
where in the last step we used that~$\phi|_{\I \setminus \U} \geq 1$ and
thus~$\tilde{\phi}|_{\I \setminus \U} \geq 1$.
Combining the above inequalities,
we conclude that~$\Sact_\I[\rho, \tilde{\phi}] \leq
\Sact_\I[\tilde{\rho}, \tilde{\phi}]$. Since~$\tilde{\rho}$ is arbitrary, the measure~$\rho$ is indeed a minimizer. 
\QED
In particular, this lemma allows us to always replace the external potential~$\phi\in\B^+(\I)$ by the
function~$\tilde{\phi}$ defined by
\[ \tilde{\phi}(x) = \left\{ \begin{array}{cl} \phi(x) & \text{if $x \in \supp(\rho)$} \\[0.2em]
\displaystyle C & \text{if $x \not\in \supp(\rho)$} \,, \end{array} \right. \]
where $C\geq\min(\sup_\I \phi, 1)$ is a constant.
Clearly, $\tilde{\phi}$ is again lower semi-continuous, because $\phi\leq1$ on $\supp(\rho)$ and the set~$\I\backslash\supp(\rho)$ is open in $\I$.
It is also worth noting that, due to the identity~\eqref{euler1}, the points of discontinuity
of~$\tilde{\phi}$ all lie on the boundary of~$\supp(\rho)$.

As a last observation before coming to our existence results, we now
explain an improvement of the positivity result~\eqref{euler2} which seems of independent
interest (although we will not need it later on). For a given minimizer~$\rho$
of the inner variational principle we introduce the set
\[ \K = (\L \rho + \phi)^{-1}(1) \subset \I \:. \]
According to~\eqref{euler1}, we know that~$\supp \rho$ is a subset of~$\K$.
The next proposition shows that the operator~$\L_\rho$ defined in~\eqref{Lrho-op} remains non-negative if we
extend it to the Hilbert space obtained by adding to $\H_\rho$ a one-dimensional subspace supported
in~$\mathscr{K} \setminus \supp \rho$ (for related results in the non-compact setting
see~\cite[Section~3.5]{lagrange}).
\begin{Prp} \label{prpEL2}
Let~$\rho \in \M^+(\I)$ be a minimizer of the inner variational principle with external potential~$\phi$.
Choosing a measure~$\nu \in \M^+(\K \setminus \supp \rho)$ 
with~$\nu \neq 0$,
we introduce the Hilbert space~$\H_\text{ext}$ by
\begin{align*}
\H_\text{ext} &= \H_\rho \oplus \R \\
\Big\la \begin{pmatrix} \psi \\ x \end{pmatrix} \!, \begin{pmatrix} \chi \\ y \end{pmatrix}
\Big\ra &= \int_\I \psi(z)\, \chi(z)\: d\rho(z) +
x y\, \nu(\I) \:,
\end{align*}
and introduce the operator~$\L_\text{ext} \in \Lin(\H_\text{ext})$ by
\[  \L_\text{ext} \begin{pmatrix} \psi \\ x \end{pmatrix} = \begin{pmatrix}
\L(\psi \rho + x \nu) \\[0.2em] \nu(\I)^{-1}\: \big\la \L(\psi \rho + x \nu), \nu \big\ra \end{pmatrix} \:. \]
Then the operator~$\L_\text{ext}$ is non-negative.
\end{Prp}
\Proof Otherwise there would be a vector~$(\psi, x) \in \H_\text{ext}$
with $\la (\psi, x) | \L_\text{ext} (\psi, x) \ra < 0$.
Possibly by flipping the sign of this vector, we can arrange that~$x \geq 0$. Then the family of measures
\[ \tilde{\rho}(t) = (1 + t \psi) \rho + t x\, \nu \qquad \text{with~$t \geq 0$} \]
is a one-parameter family of measures in~$\M^+(\I)$. A short calculation shows that
\[ \frac{d}{dt} \Sact[\tilde{\rho}(t), \phi]|_{t=0} = 0 \:,\qquad
\frac{d^2}{dt^2} \Sact[\tilde{\rho}(t), \phi]|_{t=0} = 2 \,\la (\psi, x), \L_\text{ext} (\psi, x) \ra < 0\:, \]
in contradiction to the minimality of~$\rho$.
\QED
The next example shows why in the previous proposition it is in general impossible
to extend~$\H_\rho$ by a two-dimensional subspace.
\begin{Example} {\em{
We let~$\I=\{1,2,3\}$, $\phi \equiv 0$, $\s=1$ and choose the Lagrangian~$\L$ as
\beq \label{Lmat}
\L = \begin{pmatrix} 1 & 1 & 1 \\ 1 & 1 & 2 \\ 1 & 2 & 1 \end{pmatrix} \:.
\eeq
The measure~$\rho$ is a weighted counting measure with weights~$(\rho_1, \rho_2, \rho_3)$.
The estimate
\[ \la \L \rho, \rho \ra = (\rho_1+\rho_2+\rho_3)^2 + 2 \rho_2 \rho_3- 2 (\rho_1 + \rho_2 + \rho_3) 
\geq (\rho_1+\rho_2+\rho_3 - 1)^2 -1 \]
shows that the measure~$\rho = (1,0,0)$ is a minimizer.
Moreover, the set~$\K$ equals~$\I$. If we extended~$\H_\rho$ by a two-dimensional space,
the operator~$\L_\text{ext}$ would not be positive semi-definite, because the matrix in~\eqref{Lmat}
has a negative eigenvalue.
\QEDrem
}} \end{Example}

\subsection{Solving the Inner Variational Principle} \label{sec-existinner}
We begin with an a-priori estimate of the total volume.
\begin{Lemma} \label{lemma-apriori}
There is a constant~$C=C(\L, \I)$ such that for every external potential~$\phi \in \B^+(\I)$
and for every~$\rho \in \M^+(\I)$ the following implication holds:
\[ \Sact_\I[\rho, \phi] \leq 0 \qquad \Longrightarrow \qquad \rho(\I) \leq C \:. \]
\end{Lemma}
\Proof As~$\I$ is compact and~$\L$ is continuous, the inequality~\eqref{posdiag} implies
that there is a parameter~$\delta>0$ such that~$\L(x,x)> 2 \delta$ for all~$x \in \I$.
Moreover, every~$x \in \I$ has an open neighborhood~$U(x)$ such that~$\L(y,z) > \delta$
for all~$y,z \in U(x)$. By compactness, $\I$ can be covered by a finite number of
such neighborhoods~$U_1, \ldots, U_N$, and the sets $V_k:=U_k \setminus (U_1 \cup \cdots \cup U_{k-1})$ still cover $\I$. Then for any measure~$\rho \in \M^+(\I)$,
\beq \label{rhoes}
\la \L \rho, \rho \ra \geq \sum_{k=1}^N \iint_{V_k \times V_k} \L(x,y)\: d\rho(x)\, d\rho(y)
\geq \delta \sum_{k=1}^N \rho(V_k)^2
\geq \frac{\delta}{N}\: \rho(\I)^2\:,
\eeq
where in the last step we applied H\"older's inequality
\[ \rho(\I) = \sum_{k=1}^N \rho(V_k) \leq \sqrt{N} \:\Big( \sum_{k=1}^N \rho(V_k)^2 \Big)^\frac{1}{2}\:. \]

The inequality~\eqref{rhoes} allows us to estimate the inner action~\eqref{SIdef} by
\[ \Sact_\I \geq \frac{\delta}{N}\: \rho(\I)^2 - 2\, \rho(\I) \:. \]
Thus if~$\rho(\I) > C:= 2N/\delta$, then the action is positive.
\QED

Using this estimate, we can show existence of minimizers of the inner variational principle.
\begin{Thm} \label{thm-exist}
For any given potential~$\phi \in \B^+(\I)$, the action~$\Sact_\I[\,.\,, \phi]$
has a minimizer~$\rho \in \M^+(\I)$.
\end{Thm}
\Proof Since~$\Sact_\I [0, \phi]=0$, it is obvious that~$s := \inf_{\tilde{\rho} \in \M^+(\I)} \Sact_\I [\tilde{\rho}, \phi] \leq 0$.
On the other hand, Lemma~\ref{lemma-apriori} and the fact that $\I$ is compact and 
that~$\L(x,y)$ is continuous imply that~$s>-\infty$. We choose a minimizing
sequence~$(\rho_n)_{n\in\N}$ with~$\rho_n \in \M^+(\I)$ and~$\Sact[\rho_n, \phi] \leq 0$ for all $n\in\N$.
According to Lemma~\ref{lemma-apriori}, the total volume of the measures~$\rho_n$ is uniformly bounded.
Hence
$$\int_\I |f(x)|\,d\rho_n(x) \leq \rho_n(\I) \cdot \sup_\I|f| \leq C \cdot \sup_\I|f|$$
for any function $f \in C^0(\I)$ and any $n\in \N$, implying that the sequence $(\rho_n)$ is bounded in $C^0(\I)^*$.
The Banach-Alaoglu theorem (see e.g. \cite[Theorem IV.21]{reed+simon}) yields a subsequence, again denoted by $(\rho_n)_{n\in\N}$, which converges to a functional~$\rho$ in the weak-$\ast$ topology on $C^0(\I)^*$. According to the Riesz representation theorem, $\rho$ is represented by a measure~$\rho\in\M^+(\I)$.
It remains to show that $\Sact_\I[\rho, \phi] = s$. The weak-$\ast$ convergence $\rho_n \rightharpoonup \rho$ immediately yields
$$\la \L \rho_n, \rho_n \ra \rightarrow \la \L \rho, \rho \ra \qquad \text{ and } \qquad \la 1, \rho_n \ra \rightarrow \la 1, \rho \ra\,, $$
since $1$ and $L(x,y)$ are continuous functions. Lower semi-continuity of $\phi$ implies that
$$\la \phi, \rho \ra \leq \liminf_{n\rightarrow\infty} \la \phi, \rho_n \ra$$
(see~\cite[Proposition 1.3.2]{ambrosio+tilli}), and hence 
$$s \leq \Sact_\I[\rho, \phi] \leq \lim_{n\rightarrow\infty}\Sact_\I[\rho_n, \phi]=s\,,$$
concluding the proof.
\QED

Combining Proposition~\ref{prp-euler} with Lemma~\ref{nlemma-supp} and Theorem~\ref{thm-exist},
we can state a sufficient criterion for a measure $\rho\in\M^+(\I)$ to be a minimizer.
\begin{Thm} \label{prp-eulersupp}
Let~$\rho$ be a solution of the EL equations~\eqref{euler1} and~\eqref{euler2}  with external potential~$\phi\in\B^+(\I)$ and assume that~$\supp(\rho) = \{x\in\I\,|\,\phi(x)\leq1\}$. Then~$\rho$ is a minimizer of the action~$\Sact_\I[\,.\,,\phi]$.
\end{Thm}
\Proof
Let $\tilde{\rho} \in \M^+(\I)$ be a minimizer of $\Sact_\I[\,.\,, \phi]$. Then $\supp(\tilde{\rho})\subseteq\{\phi\leq1\}=\supp(\rho)$ according to Lemma~\ref{nlemma-supp}. Moreover, since $\rho$ and $\tilde{\rho}$ are solutions of the EL equation~\eqref{euler1}, we know that
\begin{align}
 \Sact_\I[\rho,\phi] &= -\la \L\rho, \rho \ra = \la \phi-1, \rho \ra \,, \label{S-rhorel} \\
 \Sact_\I[\tilde{\rho},\phi] &= -\la \L\tilde{\rho}, \tilde{\rho} \ra = \la \phi-1, \tilde{\rho} \ra \,, \label{S-rhotilrel} \\
 \L\tilde{\rho} &= \L\rho \quad \text{ on }\supp(\tilde{\rho})\,. \label{L-rhorhotilrel} 
\end{align}
Now consider the convex combination~$\rho_\tau:=\tau\tilde{\rho}+(1-\tau)\rho\in\M^+(\I)$ for $\tau\in[0,1]$. Using the identities~\eqref{S-rhorel}--\eqref{L-rhorhotilrel}, the $\tau$-derivative of the action of $\rho_\tau$ is
computed by
\begin{align}
	\nonumber \frac{d}{d\tau} \Sact_\I[\rho_\tau,\phi] &= 2\tau \la \L\tilde{\rho}, \tilde{\rho} \ra + (2\tau-2) \la \L\rho, \rho \ra +(2-4\tau) \la \L\rho, \tilde{\rho} \ra + 2 \la \phi-1, \tilde{\rho} \ra - 2 \la \phi-1, \rho \ra \\
	\nonumber &= 2\tau \la \L\tilde{\rho}, \tilde{\rho} \ra + (2\tau-2) \la \L\rho, \rho \ra +(2-4\tau) \la \L\rho, \tilde{\rho} \ra - 2 \la \L \rho, \tilde{\rho} \ra + 2 \la \L\rho, \rho \ra \\
  &= 2\tau \la \L(\tilde{\rho}-\rho), \tilde{\rho}-\rho \ra \,.	\label{Stau-deriv}
\end{align}
Since $\tilde{\rho}-\rho \in \M(\supp(\rho))$, the EL equation~\eqref{euler2} implies that the last line in~\eqref{Stau-deriv} is non-negative. We conclude that~$\Sact_\I[\rho,\phi]\leq\Sact_\I[\tilde{\rho},\phi]$, and thus~$\rho$ is a minimizer.
\QED

\subsection{Solving the Initial Value Problem} \label{sec-existinit}
As explained in Section~\ref{sec-initdata}, the role of the external potential is to ensure that solutions of the inner variational principle satisfy the constraints imposed by the initial data.
We now analyze for which initial data it is possible to find such an external potential.
\begin{Def} \label{defadmissible}
The initial data~$\rho_0$ (as in Definition~\ref{defcauchy}) or~$(\rho_0, \I_0)$ (as in Definition~\ref{defcauchy2})
is called {\bf{admissible}} if there exists an external potential~$\phi \in \B^+(\I)$
and a measure~$\rho \in \M^+(\I)$ which is a solution of the corresponding initial value
problem.
\end{Def}
In the setting of Definition~\ref{defcauchy}, the admissible initial data is characterized by the following lemma.
\begin{Lemma} \label{lemmaadm1}
The initial data~$\rho_0 \in \M^+(\I)$ is admissible if and only if the following two conditions hold:
\begin{align}
\L \rho_0|_{\supp \rho_0} &\leq 1 \label{adm1} \\
\la \L \mu, \mu \ra &\geq 0 \qquad \text{for all~$\mu \in \M(\supp \rho_0)$}\:. \label{adm2}
\end{align}
\end{Lemma}
\Proof Suppose that~$\rho$ is a solution of the initial value problem with external potential~$\phi$.
Since~$\rho \geq \rho_0$, we know that~$\L \rho \geq \L \rho_0$. Hence the
EL equation~\eqref{euler1} can be satisfied only if the condition~\eqref{adm1} holds.
Moreover, combining the EL equation~\eqref{euler2} with the fact
that~$\supp \rho_0 \subset \supp \rho$, one sees that also the condition~\eqref{adm2}
is necessary.

In order to prove that these conditions are also sufficient, assume that a measure~$\rho_0$
satisfies~\eqref{adm1} and~\eqref{adm2}. We set
\beq \label{phiexist}
\phi(x) = \left\{ \begin{array}{cl} 1- (\L \rho_0)(x) & \text{if $x \in \supp(\rho_0)$} \\[0.2em]
2 & \text{if $x \not\in \supp(\rho_0)$}\:. \end{array} \right.
\eeq
Let us verify that~$\rho_0$ is a minimizer of the inner variational principle with external potential~$\phi$.
To this end, let~$\rho$ be a minimizer. Then Lemma~\ref{nlemma-supp} yields that~$\supp \rho \subset \supp \rho_0$.
Thus setting~$\mu = \rho - \rho_0 \in \M(\supp \rho_0)$, we
may apply~\eqref{adm2} to obtain
\begin{align*}
\Sact_\I[\rho, \phi] &= \la \L \rho, \rho \ra + 2\, \big\la (\phi-1_\I), \rho \big\ra \\
&= \la \L (\rho_0 + \mu), \rho_0+\mu \ra + 2\, \big\la \phi-1, \rho_0+\mu \big\ra \\
&= \Sact_\I[\rho_0, \phi] + 2 \,\la \L \rho_0, \mu \ra + \big\la \L \mu, \mu \ra + 2\, \la \phi-1, \mu \big\ra \\
&\geq \Sact_\I[\rho_0, \phi] + 2\, \big\la \L \rho_0 + \phi-1, \mu \big\ra
= \Sact_\I[\rho_0, \phi] \:,
\end{align*}
where in the last step we applied~\eqref{phiexist}.
\QED
We now extend the previous result to the setting
of Definition~\ref{defcauchy2}.
\begin{Lemma} \label{lemmaadm2}
The initial data~$(\rho_0, \I_0)$ is admissible if and only if the following two conditions hold:
\begin{align}
\L \rho_0|_{\I_0 \cup \supp \rho_0} &\leq 1 \label{adm21} \\
\la \L \mu, \mu \ra &\geq 0 \qquad \text{for all~$\mu \in \M(\I_0 \cup \supp \rho_0)$}\:. \label{adm22}
\end{align}
\end{Lemma}
\Proof Combining the EL equations~\eqref{euler1} and~\eqref{euler2} with
the conditions~(b) and~(c) in Definition~\ref{defcauchy2}, it is obvious that the conditions~\eqref{adm21}
and~\eqref{adm22} are necessary. In order to show that they are also sufficient,
assume that~$\rho_0$ is a measure with the above properties.
We let~$\nu \in \M^+(\I)$ be a measure with~$\supp \nu = \I_0$. By rescaling~$\nu$ we can arrange
that~$\sup_{\I_0 \cup \supp \rho_0} (\L \nu) \leq 1$. We introduce the series of measures~$(\rho_0^n)$ by
\beq \label{rhondef}
\rho_0^n = \left(1-\frac{1}{n} \right) \rho_0 + \frac{\nu}{n} \:\in\: \M^+(\I_0)\:.
\eeq
These measures have the property that~$\supp \rho_0^n = \I_0 \cup \supp \rho_0$.
Moreover, they satisfy the assumptions of Lemma~\ref{lemmaadm1}. Thus,~$\rho_0^n$ is a minimizer
of~$\Sact_\I$ with external potential~$\phi_n$ of the form~\eqref{phiexist}.
It is obvious from the construction that the conditions~(a)-(c) in Definition~\ref{defcauchy2}
are satisfied.

Taking the limit~$n \rightarrow \infty$, we conclude from~\eqref{rhondef} and~\eqref{phiexist}
that~$\rho_0^n \rightarrow \rho_0$ and~$\phi_n$ converges uniformly to~$\phi \in \B^+(\I)$.
It follows by continuity that~$\rho_0$ is again a minimizer of~$\Sact_\I$ with external potential~$\phi$.
Moreover, continuity yields that the conditions~(a)-(c) in Definition~\ref{defcauchy2} are preserved
in the limit.
\QED

A special class of admissible initial data is given by the following subsets of~$\I$:
\begin{Def} A subset~$\I_0 \subset \I$ is called {\bf{totally space-like}}
if~$\L(x,y)=0$ for all~$x,y \in \I_0$ with~$x \neq y$.
\end{Def} \noindent
Note that the continuity argument in the proof of Lemma~\ref{lemma-apriori} shows immediately
that every totally space-like set is discrete.
\begin{Lemma} Choosing~$\rho_0=0$ and~$\I_0 \subset \I$ as a totally space-like subset,
the initial data~$(\rho_0, \I_0)$ is admissible.
\end{Lemma}
\Proof The condition~\eqref{adm21} is trivially satisfied. Using that~$\I_0$ is totally space-like,
the expression in~\eqref{adm22} simplifies to
\[ \la \L \mu, \mu \ra = \sum_{x \in \I_0} \L(x,x)\, \mu \big( \{x\} \big)^2 \:, \]
which is obviously non-negative.
\QED

\subsection{Existence of Optimal Solutions} \label{sec-existopt}
In Section~\ref{sec-optpot} we introduced several notions of an optimal solution of the initial value problem.
We now prove that such optimal solutions exist, provided that the initial data is admissible.
\begin{Thm}\label{thm-exopt} Assume that the initial data~$\rho_0$ or~$(\rho_0, \I_0)$ is admissible.
Then there exist solutions of the optimization problems~(A), (B) and (C) in~\eqref{optimize1}, and (D) in~\eqref{optimize2}.
\end{Thm}
\Proof 
We first consider the problems~(A), (B) and~(C). In each case, we can choose a minimizing or maximizing sequence~$((\rho_n, \phi_n))_{n\in\N} \subset \mathfrak{S}_\I(\rho_0,\I_0)$, where $\mathfrak{S}_\I$ is the solution set defined in~\eqref{solset}.
In view of Lemma~\ref{nlemmareplace}, we may replace the functions~$\phi_n$
by the functions
\beq \tilde{\phi}_n(x) = \left\{ \begin{array}{cl} \phi_n(x) & \text{if $x \in \supp(\rho_n)$} \\[0.2em]
\displaystyle 2 & \text{if $x \notin \supp(\rho_n)$} \end{array} \right. \label{phioptrepl} \eeq
(note that this replacement leaves the functionals in~(A), (B) and~(C) unchanged).
Since each $\rho_n$ is a minimizer of $\Sact_\I[\,.\,, \tilde{\phi}_n]$, Lemma~\ref{lemma-apriori} implies that the volume is uniformly bounded, i.e. there is a constant $C_V>0$ such that
\beq \rho_n(\I) \leq C_V \quad \text{ for all $n\in\N$.} \label{vol-bound} \eeq
Thus in case (C), the sequence $\rho_n(\I)$ converges to a value $M_C\in[0,C_V]$.
From the definition of~$\tilde{\phi}_n$ and the equation~\eqref{euler1}, we know that
$$ 0 \leq \tilde{\phi}_n|_{\supp(\rho_n)} \leq 1 \quad \text{ for all $n\in\N$}\,. $$
Thus in case (B), the sequence $\max_{\supp(\rho_n)}\tilde{\phi}_n$ converges to a value $M_B\in[0,1]$.
Combining~\eqref{vol-bound} with Corollary~\ref{cor-minact}, we see that the action is bounded from below,
$$ \Sact_\I[\rho_n,\tilde{\phi}_n] = - \la \L \rho_n, \rho_n \ra \geq - C_V \cdot \sup_{\I\times\I}\L(x,y) =: -C_S \quad \text{ for all $n\in\N$.} $$
Thus in case (A), the sequence $\Sact_\I[\rho_n,\tilde{\phi}_n]$ converges to a value $M_A\in[-C_S,0]$.

The inequality~\eqref{vol-bound} also implies that the sequence $(\rho_n)$ is bounded in $C^0(\I)^*$. 
Thus the Banach-Alaoglu theorem yields a subsequence, again denoted by $(\rho_n)_{n\in\N}$, which converges to a functional~$\rho$ in the weak-$\ast$ topology on $C^0(\I)^*$. According to the Riesz representation theorem, $\rho$ is represented by a measure~$\rho\in\M^+(\I)$.
Since the constant function $f\equiv1$ is continuous on $\I$, the weak-$\ast$ convergence $\rho_{n}\rightharpoonup\rho$ implies that the volume converges,
\[ \rho_{n}(\I) = \int_\I 1\,d\rho_n \stackrel{n\rightarrow\infty}{\longrightarrow} \int_\I 1\,d\rho = \rho(\I) \,. \]
Next, we introduce the function~$\phi:\I\rightarrow\R$ by
\beq \phi(x) = \left\{ \begin{array}{cl} 1- (\L \rho)(x) & \text{if $x \in \supp(\rho)$} \\[0.2em]
\displaystyle 2 & \text{if $x \not \in \supp(\rho)$}\:. \end{array} \right. \label{def-phiopt} \eeq
Since $\text{dist}(x,\supp(\rho_n))\rightarrow 0$ for any $x\in\supp(\rho)$ and since $1-\L\rho_n\geq 0$ on $\supp(\rho_n)$, we conclude from the pointwise convergence $1-L\rho_n\rightarrow 1-\L\rho$ that 
$0\leq\phi\leq2$, which implies that~$\phi\in\B^+(\I)$.
Moreover, the supremum of the external potential converges,
$$\max_{\supp(\rho_n)}\tilde{\phi}_n \stackrel{n\rightarrow\infty}{\longrightarrow} \max_{\supp(\rho)}\phi \,.$$
It is obvious from~\eqref{def-phiopt} that $\rho$ is a solution of the EL equation~\eqref{euler1} with external potential~$\phi$.
Combining this fact with the continuity of $\L(x,y)$ we find that the action converges,
\[ \Sact_\I[\rho_n, \tilde{\phi}_n] = - \la \L \rho_n, \rho_n \ra \stackrel{n\rightarrow\infty}{\longrightarrow} - \la \L \rho, \rho \ra = \Sact_\I[\rho,\phi]\,. \]
The weak-$\ast$ convergence $\rho_n\rightharpoonup\rho$ and the continuity of~$\L(x,y)$ also imply that the EL equation~\eqref{euler2} holds on~$\supp(\rho)$. Therefore, we can apply Proposition~\ref{prp-eulersupp} and see that~$\rho$ is a minimizer of~$\Sact_\I[\,.\,,\phi]$.
Finally, the condition~$\rho \geq \rho_0$ (in the setting of Definition~\ref{defcauchy})
or the conditions~(a)-(c) (in the setting of Definition~\ref{defcauchy2}) are obviously preserved in the
limit~$n \rightarrow \infty$. This concludes the proof for the optimization problems~(A), (B) and (C). 

Considering the optimization problem (D), we know from case (C) that there exist elements in $\mathfrak{S}_\I(\rho_0,\I_0)$ which maximize the volume, i.e.\ the set~$\mathfrak{S}_{\I}^\text{maxV}(\rho_0,\I_0)$ is non-empty.
Choosing a maximizing sequence~$(\rho_n,\phi_n)_{n\in\N}\subset\mathfrak{S}_{\I}^\text{maxV}(\rho_0,\I_0)$ for the action $\Sact_\I$, we can use the same arguments as in case (A) to see that there is a pair $(\rho,\phi)\in\mathfrak{S}_{\I}^\text{maxV}(\rho_0,\I_0)$ with maximal action.
\QED

\section{Uniqueness Results} \label{sec-uniqueness}
Having settled the existence problem, our next task is to analyze the uniqueness of solutions. More precisely, the first question which we shall address in this section is on which subsystems of~$\I$ the solution of the initial
value problem is uniquely determined for any choice of the external potential. The second question concerns the freedom in choosing the external potential~$\phi$, and whether this freedom can be removed by
working with optimal solutions as introduced in Section~\ref{sec-optpot}. 

\subsection{The Domain of Dependence} \label{sec-domdep}
In this section, we shall investigate on which subsystems of $\I$ the solution of the initial value problem is unique and whether there is a ``largest'' subsystem having this property. We consider given initial data~$(\rho_0,\I_0)$ with~$\rho_0 \in \M^+(\I)$ and a (possibly empty) closed subset~$\I_0 \subset \I$.
\begin{Def}
A Borel subset~$\Bm\subset\I$ {\bf{encloses the initial data}} 
if~$\supp \rho_0 \cup \I_0 \subset \Bm$.
\end{Def}
A sufficient criterion for uniqueness on the subsystem~$\Bm\subset\I$ is that the Lagrangian is positive definite in the following sense.
\begin{Prp} \label{prop-uniquecrit}
Let~$\Bm \subset \I$ a set which encloses the initial data. If the Lagrangian is {\bf{positive definite}} on~$\Bm$ in the sense that~$\la \L \mu, \mu \ra > 0$ for every
non-zero signed measure~$\mu \in \M(\Bm)$ with~$\L \mu|_{\supp \rho_0 \cup \I_0} \equiv 0$, 
then for any given external potential~$\phi\in\B^+(\Bm)$, there is at most one solution of the corresponding initial value problem in~$\Bm$.
\end{Prp}
\Proof Assume that there is a external potential $\phi \in \B^+(\Bm)$ for which the
initial value problem in~$\Bm$ has two distinct solutions~$\rho, \tilde{\rho} \in \M^+(\Bm)$. Then~$\mu := \tilde{\rho} - \rho \in \M(\Bm)$ is non-zero, and the EL equations for~$\rho$ and~$\tilde{\rho}$ imply
that~$\L \mu|_{\supp \rho_0 \cup \I_0} \equiv 0$. Thus the following inequality holds,
\beq \label{munull} 0 < \la \L \mu, \mu \ra = \la \L \tilde{\rho}, \tilde{\rho} \ra  + \la \L \rho, \rho \ra  - 2 \la \L \tilde{\rho}, \rho \ra \,. \eeq
Defining the measure~$\hat{\rho}:=\frac{1}{2}(\tilde{\rho}+\rho)\in\M^+(\Bm)$, it follows from \eqref{munull} that
\begin{align*}
	\Sact_\Bm[\hat{\rho},\phi] &= \frac{1}{4}\, \la \L \tilde{\rho}, \tilde{\rho} \ra + \frac{1}{4}\, \la \L \rho, \rho \ra + \frac{1}{2}\, \la \L \tilde{\rho}, \rho \ra + \la \phi-1, \tilde{\rho} + \rho \ra \\
	&< \frac{1}{2}\, \la \L \tilde{\rho}, \tilde{\rho} \ra + \frac{1}{2}\, \la \L \rho, \rho \ra + \la \phi-1, \tilde{\rho} + \rho \ra = \frac{1}{2}\, \Sact_\Bm[\tilde{\rho},\phi] + \frac{1}{2}\, \Sact_\Bm[\rho,\phi]\,.
\end{align*}
This is a contradiction because~$\rho$ and~$\tilde{\rho}$ are both minimizers.
\QED

Unfortunately, the uniqueness is in general not preserved when taking
unions of closed sets, as the following example shows.
\begin{Example} {\bf{(The heat kernel on the unit circle)}} \label{ex-circle}
{\em{	
We consider the sphere~$\I=S^1=\R \!\!\mod 2\pi$ with initial data~$\rho_0=0$
and denote the Haar measure on~$S^1$ by~$dx$. The real Hilbert space~$L^2(\I,dx)$ has an orthonormal basis consisting of the constant function~$1$ and the functions~$\cos(kx)$ and~$\sin(kx)$,
where~$x\in[0,2\pi]$ and~$k\in\N$. The heat kernel
$$\L(x,y)=\frac{1}{2\pi}+\frac{1}{\pi}\sum_{k=1}^\infty e^{-k^2} \cos \!\big(k(x-y) \big) $$
is the integral kernel of a positive definite compact operator on~$L^2(\I,dx)$. Approximating a signed Borel measure~$\mu \in \M(\I)$ in the weak-$\ast$ topology by functions~$\psi_n\in L^2(\I,dx)$, we see that the Lagrangian~$\L(x,y)$ is positive definite on~$\I$ in the sense of Proposition~\ref{prop-uniquecrit}. Hence
for every external potential~$\phi\in\B^+(\I)$, the inner variational principle has a unique minimizer.
For example, choosing~$\phi\equiv C$ as a constant function,
a short computation shows that the measure~$\rho=(1-C)\,dx$ is the unique
minimizer of~$\Sact_\I[\,.\,,\phi]$.

Now we modify the Lagrangian as follows,
\beq \label{Lmod}
\tilde{\L}(x,y)=\L(x,y)-\frac{1}{\pi e} \cos(x-y) \,.
\eeq
Again considering~$\L$ as the integral kernel of an operator on~$L^2(\I, dx)$,
the resulting operator is only positive semi-definite. It has a two-dimensional kernel
spanned by the functions~$\sin x$ and~$\cos x$.
Again approximating a measure~$\mu \in \M(\I)$, we sees that the Lagrangian is
still positive semi-definite in the sense that~$\la \mu, \tilde{\L}\mu \ra \geq 0$ for
any~$\mu\in\M(\I)$. But the fact that the Lagrangian is no longer positive definite
implies that uniqueness is lost. For example, choosing the external potential~$\phi\equiv0$, the initial value problem in~$\I$ has a 2-parameter family of solutions, given by
$$d\rho = \big( 1 + \alpha \,\cos(x)+\beta \,\sin(x) \big) \:dx \quad \text{with}
\quad 0\leq |\alpha|+|\beta| \leq1\,.$$

We next consider a proper closed subset~$\J \subset \I$ of the unit circle.
The following argument shows that the Lagrangian~$\tilde{\L}$ is positive definite on~$\M(\J)$:
Assume conversely that there is a non-trivial~$\mu \in \M(\J)$ with~$\la \mu, \tilde{\L}\mu \ra = 0$.
Extending~$\mu$ by zero to a measure in~$\M(\I)$, the resulting measure is not
the Haar measure on~$S^1$. Hence there is a function~$\psi \in C^0(\I)$
with~$\int_{I} \psi (d\mu - dx) \neq 0$. Since the trigonometric functions are dense in~$C^0(\I)$,
we conclude that there is~$k>1$ such that
\beq \label{nonvanish}
\int_\J \cos(kx)\: d\mu(x) \neq 0 \qquad \text{or} \qquad
\int_\J \sin(kx)\: d\mu(x) \neq 0 \:.
\eeq
Using the representation of the Lagrangian in term of trigonometric functions, we obtain
with the sum rules
\[ \la \mu, \tilde{\L}\mu \ra =
\frac{1}{2\pi}\: \mu(\J)^2 +\frac{1}{\pi} \sum_{k=2}^\infty e^{-k^2} 
\left( \Big( \int_{\J} \cos(kx)\, d\mu \Big)^2 + \Big( \int_{\J} \sin(kx)\, d\mu \Big)^2 \right) . \]
This is strictly positive by~\eqref{nonvanish}, a contradiction.

The positivity of~$\tilde{L}$ on~$\J$ implies 
in view of Proposition~\ref{prop-uniquecrit} that the initial value problem in the set~$\J$
has at most one solution.
Considering the initial value in the sets
$$\J_n = \Big\{ e^{i\varphi}\in S^1\ \Big|\, \varphi\in \Big[\frac{1}{n},2\pi-\frac{1}{n}\Big] \Big\} \:, $$
we conclude that for every~$n$ and every external potential~$\phi\in\B^+(\J_n)$,
the solution of the initial value problem is unique.
However, on the set~$\I = \overline{\cup_n \J_n}$, the initial value problem
does in general {\em{not}} have a unique solution.
}}
\QEDrem
\end{Example}

Our method for bypassing this loss of uniqueness is to work instead of closed sets
with \textit{open} subsets~$\Omega\subset\I$ and to consider
minimizers which are supported away from the boundary:
\begin{Def} 
Let $\Omega\subset \I$ be an open set which encloses the initial data.
\begin{itemize} 
\item[(i)]  A solution~$(\rho,\phi)$ of the initial value problem in~$\overline{\Omega}$
(with~$\phi \in \B^+(\overline{\Omega})$) is called
{\bf{interior solution}} if $\supp(\rho) \subset \Omega$.
\item[(ii)] If for every external potential~$\phi \in \B^+(\overline{\Omega})$, there is at most one interior solution of the corresponding initial value problem in~$\overline{\Omega}$, then~$\Omega$ is called {\bf{dependent}}.
\end{itemize}
\end{Def} \noindent
The next lemma gives a simple but useful property of dependent sets.
\begin{Lemma} If~$\Omega$ is dependent, so is every subset~$\Omega' \subset \Omega$
which encloses the initial data.
\end{Lemma}
\Proof Suppose that~$(\rho,\phi)$ is an interior solution of the initial value problem
in~$\overline{\Omega'}$. Then according to Lemma~\ref{nlemma-supp},
$(\rho,\tilde{\phi})$ is an interior solution of the initial value problem in~$\overline{\Omega}$, where the
external potential~$\tilde{\phi}$ is given by
\[ \tilde{\phi}(x) =  \left\{ \begin{array}{cl} \phi(x) & \text{if $x \in \overline{\Omega'}$} \\[0.2em]
2 & \text{if $x \in \overline{\Omega} \setminus \overline{\Omega'}$}\:. \end{array} \right. \]
Thus the uniqueness of interior solutions in~$\overline{\Omega}$ implies uniqueness in~$\overline{\Omega'}$.
\QED
The uniqueness criterion in Proposition~\ref{prop-uniquecrit} can be reformulated in a straightforward
way to obtain a sufficient criterion for dependent sets.
\begin{Prp} \label{prop-dependcrit}
Let~$\Omega \subset \I$ be a set which encloses the initial data. If~$\la \L \mu, \mu \ra > 0$ for every
non-trivial signed measure~$\mu \in \M(\Omega)$ with~$\L \mu|_{\supp \rho_0 \cup \I_0} \equiv 0$,
then~$\Omega$ is dependent.
\end{Prp} \noindent
The notion of dependent sets is preserved when taking unions, making it
possible to construct maximal sets, as we now explain.
\begin{Def} A dependent subset~$\Omega \subset \I$
is called {\bf{maximally dependent}} if it is not the proper subset of another dependent set.
\end{Def} \noindent
\begin{Prp} If for given initial data~$(\rho_0, \I_0)$ there is a dependent set,
then there is a maximally dependent set.
\end{Prp}
\Proof 
This follows from a standard argument using Zorn's Lemma. Namely, on the set of dependent subsets of~$\I$,
we consider the partial order given by the inclusion of sets. By separability of~$\I$ , we can restrict attention to countable chains in this partially ordered set. Let $\Omega_1\subset\Omega_2\subset\Omega_3\subset\ldots$ be such a chain and define~$\Omega=\cup_{n\in\N}\Omega_n$. Then~$\Omega$ is certainly open and 
encloses the initial data. It remains to show that it is again dependent.
Thus, for any~$\phi\in\B^+(\overline{\Omega})$ we
let~$(\rho,\phi)$ and~$(\tilde{\rho},\phi)$ be two interior solutions of the initial value problem in~$\overline{\Omega}$, i.e.\ $\supp(\rho),\supp(\tilde{\rho})\subset\Omega$. Then every~$x\in\supp(\rho)$
has an open neighborhood contained in~$\subset\Omega_n$. Since $\supp(\rho)$ is compact, we can cover it by a finite number of such neighborhoods, implying that there is~$N\in\N$ with $\supp(\rho)\subset\Omega_N$.
By increasing~$N$, we can arrange similarly that also~$\supp(\tilde{\rho})\subset\Omega_N$.
Since $\Omega_N$ is dependent, we conclude that $\rho=\tilde{\rho}$.
\QED

We point out that there may be more than one maximally dependent subset of~$\I$.
Since we want the domain of dependence to be unique and invariantly characterized, the following
definition seems natural.
\begin{Def} For given initial data~$(\rho_0, \I_0)$ in~$\I$, we define the
{\bf{domain of dependence}} $\D(\rho_0, \I_0)$ as the intersection of all maximally dependent sets,
\[ \D(\rho_0, \I_0) = \bigcap \;\big\{ \Omega \subset \I \:\big|\: \text{$\Omega$ is maximally dependent} \big\} \:. \]
\end{Def}\noindent
By construction, it is clear that~$\supp(\rho_0)\cup\I_0 \subset \D(\rho_0, \I_0)$. However, we point out that,
since the above intersection may be uncountable, the domain of dependence need not be Borel-measurable.
Hence the initial value problem in~$\D(\rho_0, \I_0)$ need not be well-defined. Nonetheless, considering the closure~$\overline{\D(\rho_0, \I_0)}$, we have uniqueness in the following sense.
\begin{Prp} For every~$\phi \in \B^+(\overline{\D(\rho_0, \I_0)})$, there is at most one
minimizer~$\rho$ of the initial value problem in~$\overline{\D(\rho_0, \I_0)}$ 
with~$\supp \rho \subset \D(\rho_0, \I_0)$.
\end{Prp}
\Proof
Since~$\D(\rho_0, \I_0)\subset\Omega$, we know that~$\overline{\D(\rho_0, \I_0)}\subset\overline{\Omega}$ for any maximally dependent set~$\Omega$. Thus the result follows immediately from the uniqueness of
interior solutions in~$\Omega$.
\QED

\begin{Example} {\em{
In the setting of Example~\ref{ex-circle}, where~$\I=S^1$, $\rho_0=0$, and~$\tilde{\L}$
the modified heat kernel~\eqref{Lmod}, 
the set $S^1\backslash\{p\}$ is maximally dependent for any~$p\in S^1$. Hence the domain of dependence is given by
\beq \D(\rho_0) = \bigcap_{p\in S^1} \big(S^1\backslash\{p\} \big) = \varnothing\,. \label{dod-example} \eeq
Since by choosing~$\rho_0=0$ we do not prescribe any non-trivial initial data, the result~\eqref{dod-example} is consistent with what one would have expected for its domain of dependence.
}}
\QEDrem
\end{Example}

\subsection{Uniqueness of Optimal Solutions} \label{sec-uniqueopt}
A shortcoming of our approach so far is that solutions of the initial value problem
depend on the choice of an external potential.
As we saw in Proposition~\ref{prop-uniquecrit}, the positivity of the Lagrangian on a subsystem~$\Bm$ ensures uniqueness of solutions for any given external potential~$\phi\in\B^+(\Bm)$. 
We will combine this fact with the existence of optimal external potentials on closed subsystems
(see Theorem~\ref{thm-exopt}) to provide a construction which uniquely determines a solution of the initial value problem with maximal volume.

\begin{Def} \label{def-definiteset}
A closed subset~$\J\subset\I$  is called {\bf{definite}} if it encloses the initial data and
if the following conditions hold:
\begin{itemize}
	\item[(i)] $\la \L \mu, \mu \ra > 0$ for any non-zero signed measure $\mu \in \M(\J)$;
	\item[(ii)] $(\L\rho)(x)\leq1$ for any~$x\in\J$ and any solution~$(\rho,\phi)$ of the initial value problem in~$\J$.
\end{itemize}
\end{Def}

\begin{Def}\label{def-strongadm}
The initial data $(\rho_0, \I_0)$ is called {\bf{strongly admissible}} if it is admissible and if
the set~$\supp(\rho_0)\cup\I_0$ is definite.
\end{Def} \noindent
In the remainder of this section, we always assume that the initial data is non-zero and strongly admissible.
We now show that the optimization problem~(D) yields a unique measure $\rho\in\M^+(\J)$,
provided that~$\J$ is definite (the example in Section~\ref{sec-causalwedge} will show that the
optimization problems~(A)-(C) yield non-unique solutions even on definite sets).

\begin{Thm} \label{thm-definiteunique}
Consider a definite subsystem~$\J\subset\I$, and let $(\rho, \phi)$ and $(\tilde{\rho}, \tilde{\phi})$ be two solutions of the optimization problem (D) in ~$\J$. Then $\rho=\tilde{\rho}$.
\end{Thm}
\Proof
Let $(\rho,\phi)$ and $(\tilde{\rho},\tilde{\phi})$ be two solutions of the optimization problem (D),
i.e. $\Sact_\J[\rho,\phi]=\Sact_\J[\tilde{\rho},\tilde{\phi}]$ is maximal in the class $\mathfrak{S}_\J^\text{maxV}(\rho_0,\I_0)$. Possibly by increasing the external potential outside the support of
the minimizing measure (cf.\ Lemma~\ref{nlemmareplace}), we can arrange that
$$\phi=\begin{cases} 1-\L\rho\quad&\text{on }\supp \rho\\ 2 \quad&\text{otherwise} \end{cases} \qquad\text{and}\qquad 
\tilde{\phi}=\begin{cases} 1-\L\tilde{\rho}\quad&\text{on }\supp \tilde{\rho}\\ 2 \quad&\text{otherwise}\,. \end{cases}$$
Obviously, the convex combination
$$\rho_\tau:=\tau\tilde{\rho}+(1-\tau)\rho$$
is again in $\M^+(\J)$ and has maximal volume for any $\tau\in[0,1]$.
By condition~(ii) in Definition~\ref{def-strongadm}, the external potentials
$$\phi_\tau:=\begin{cases} 1-\L\rho_\tau\quad&\text{on }\supp \rho_\tau \\ 2 \quad&\text{else} \end{cases}$$
are in $\B^+(\D(\rho_0, \I_0))$ for any $\tau\in[0,1]$. 

Let $\nu_\tau\in \M^+(\J)$ be a minimizer of $\Sact_{\J}[\,.\,,\phi_\tau]$. 
Then $\supp(\nu_\tau)\subseteq\{\phi_\tau\leq1\}\subseteq\supp(\rho_\tau)$. 
The EL-equation~\eqref{euler1} yields that 
$$\L\nu_\tau=\L\rho_\tau\text{ on } \supp(\nu_\tau) \quad\text{ and }\quad \L\nu_\tau\geq\L\rho_\tau\text{ on } \supp(\rho_\tau)\backslash\supp(\nu_\tau)\,.$$
Thus, we obtain
$$\la \L (\rho_\tau-\nu_\tau), \rho_\tau-\nu_\tau \ra = \la \L(\rho_\tau-\nu_\tau), \rho_\tau \ra \leq 0\,.$$
It now follows from condition~(i) in Definition~\ref{def-strongadm} that $\nu_\tau=\rho_\tau$. Since moreover $\rho_\tau\geq\rho_0$, we conclude that $\rho_\tau$ is a solution of the initial value problem in $\J$ with external potential $\phi_\tau$. Therefore, we have $(\rho_\tau,\phi_\tau)\in\mathfrak{S}_\J^\text{maxV}(\rho_0,\I_0)$ for all $\tau\in[0,1]$.

Now assume that $\tilde{\rho}-\rho\neq0$. Then the identity
$$\la \L \rho, \rho \ra = -\Sact_\J[\rho,\phi] = -\Sact_\J[\tilde{\rho},\tilde{\phi}] =  \la \L \tilde{\rho}, \tilde{\rho} \ra$$
and the inequality
$$\frac{d^2}{d\tau^2} \la \L \rho_\tau, \rho_\tau \ra = 2 \la \L(\tilde{\rho}-\rho), \tilde{\rho}-\rho \ra > 0 \quad\text{ for all }\tau\in(0,1)\,,$$
obtained from condition (i) in Definition~\ref{def-strongadm}, yield that
$$\Sact_\J[\rho_\tau,\phi_\tau] = -\la \L \rho_\tau, \rho_\tau \ra > -\la \L \rho, \rho \ra = \Sact_\J[\rho,\phi]\,. $$
Since the pair $(\rho,\phi)$ maximizes the action in $\mathfrak{S}_\J^\text{maxV}(\rho_0,\I_0)$, this is a contradiction.
\QED

We can construct a unique solution of the initial value problem which is characterized by a certain maximality condition on the volume of~$\I$, if we consider only definite sets with the following monotonicity property.
\begin{Def} \label{def-solgerm}
Let~$\J\subset\I$ be a definite set and $(\rho,\phi)$ its optimal solution (note that~$\rho$ is unique by
Theorem~\ref{thm-definiteunique}).
Then the pair~$(\J,\rho)$ is called a {\bf{solution germ}} if for any other definite set~$\tilde{\J}\subset\I$ with optimal solution~$(\tilde{\rho},\tilde{\phi})$ the following implication holds,
\beq \tilde{\rho}(\tilde{\J}) \geq \rho(\J) \quad \Longrightarrow \quad \tilde{\rho} \geq \rho \label{solgerm} \eeq
(in the last inequality, we extend both~$\rho$ and~$\tilde{\rho}$ by zero to measures in~$\M^+(\I)$).
\end{Def}

The set
$$ {\mathcal{V}}(\rho_0,\I_0) = \big\{ V\geq0 \,\big|\,\text{there is a solution germ $(\J,\rho)$ with $\rho(\J)=V$} \big\}\subset \R^+_0 $$
is bounded in view of the a-priori estimate in Lemma~\ref{lemma-apriori}. Since~$(\supp(\rho_0),\rho_0)$ is
a solution germ, the set~${\mathcal{V}}(\rho_0,\I_0)$ is non-empty. 
The property~\eqref{solgerm} implies that for every~$V\in{\mathcal{V}}(\rho_0,\I_0)$, there is a unique solution germ~$(\J,\rho)$ with~$\rho(\J)=V$. Hence we can identify the set of all solution germs with the totally ordered set~${\mathcal{V}}(\rho_0,\I_0)\subset\R$. 
Since the set~${\mathcal{V}}(\rho_0,\I_0)$ need not be closed, there may not exist a solution germ with maximal volume. However, the next theorem shows that there is a unique limit of monotone increasing and volume-maximizing sequences of solution germs~$(\J_n,\rho_n)$, by which we mean that
\[ \rho_n\leq\rho_m \quad\text{as measures in }\M^+(\I) \text{ for all }n\leq m \]
and
\[ \lim_{n\rightarrow\infty} \rho_n(\J_n) = \sup{\mathcal{V}}(\rho_0,\I_0) \,. \]
\begin{Thm} \label{lemma-uniquerho}
There is a unique measure~$\rho\in\M^+(\I)$ that arises as the weak-$\ast$ limit of monotone increasing and volume-maximizing sequences of solution germs~$(\J_n,\rho_n)$, i.e.
$$\rho = \text{w-$\ast$-}\lim_{n\rightarrow\infty} \rho_n \,.$$
\end{Thm} \noindent
We refer to~$\rho$ as the {\bf{maximal optimal solution}}.
\Proof
Existence again follows from the Banach-Alaoglu theorem. Assume that~$(\J_n,\rho_n)$ and~$(\tilde{\J}_m,\tilde{\rho}_m)$ are two monotone increasing and volume-maximizing sequences of solution germs, such that
$$\text{w-$\ast$-}\lim_{n\rightarrow\infty}\rho_n = \rho\in\M^+(\I) \qquad\text{and}\qquad \text{w-$\ast$-}\lim_{m\rightarrow\infty}\tilde{\rho}_m = \tilde{\rho}\in\M^+(\I) \,.$$
Then we obviously have~$\rho(\I)=\sup{\mathcal{V}}(\rho_0,\I_0)=\tilde{\rho}(\I)$.
Moreover, monotonicity and weak-$\ast$ convergence imply that~$\rho\geq\rho_n$ and~$\tilde{\rho}\geq\tilde{\rho}_m$ for all~$n,m\in\N$. 
We can clearly choose subsequences~$\rho_{n_k}$ and~$\tilde{\rho}_{m_k}$, such that either~$\rho_{n_k}(\I)\leq\tilde{\rho}_{m_k}(\I)$ or~$\rho_{n_k}(\I)\geq\tilde{\rho}_{m_k}(\I)$ for all~$k\in\N$. The implication~\eqref{solgerm} then yields~$\rho_{n_k}\leq\tilde{\rho}_{m_k}$ or~$\rho_{n_k}\geq\tilde{\rho}_{m_k}$ for all~$k\in\N$. Since both subsequences converge, we conclude that~$\rho\leq\tilde{\rho}$ or~$\rho\geq\tilde{\rho}$. Now the identity~$\rho(\I)=\tilde{\rho}(\I)$ implies that~$\rho=\tilde{\rho}$.
\QED

We finally explain in which sense the maximal optimal solution is a solution of the initial value problem.
\begin{Prp}
There exists an external potential~$\phi\in\B^+(\I)$ such that~$(\rho,\phi)$ is a solution of the initial value problem in~$\I$.
\end{Prp}
\Proof
We can use the same techniques as in the proof of Theorem~\ref{thm-exopt}. Namely,
we choose a sequence of solution germs~$(\J_n,\rho_n)$, such that~$\rho=\text{w-$\ast$-}\lim\rho_n$. Let $\phi_n\in\B^+(\J_n)$ be corresponding external potentials, such that~$(\rho_n,\phi_n)$ is a solution of the initial value problem in~$\J_n$. If we replace~$\phi_n$ by the external potential~$\tilde{\phi}_n\in\B^+(\I)$ as given by~\eqref{phioptrepl}, then~$(\rho_n,\tilde{\phi}_n)$ is a solution of the initial value problem in~$\I$. Defining~$\phi\in\B^+(\I)$ by~\eqref{def-phiopt}, weak-$\ast$ convergence implies that~$\rho$ is a solution of the EL equations~\eqref{euler1} and~\eqref{euler2} with corresponding external potential~$\phi$. Proposition~\ref{prp-eulersupp} then yields that~$\rho$ is a minimizer of~$\Sact_\I[\,.\,,\phi]$, and continuity yields that~$\rho$ is a solution of the initial value problem.
\QED

\section{Examples} \label{sec-examples}
\subsection{A Constant Lagrangian} \label{sec-constlagr}
Let us analyze the simple example when the Lagrangian is constant,
$$\L(x,y)=1 \quad \text{for all }x,y\in\I\,.$$
The estimate
$$\Sact_\I[\rho,\phi] = \rho(\I)^2-2\rho(\I)+2\int_\I\phi\,d\rho \geq \rho(\I)(\rho(\I)-2)+2\rho(\I) \inf_\I \phi$$
shows that for any external potential $\phi\in\B^+(\I)$ with $\inf_\I\phi\leq1$, the minimizer of the action $\Sact_\I[\,.\,,\phi]$ is supported in the set
$$M_\phi:=\{ x\in\I \,|\, \phi(x)=\inf_\I \phi \}\,.$$
This observation 
simplifies our problem considerably, because the volume~$\rho(\I)$ is the only parameter that remains
to be varied. It follows that any measure $\rho\in\M^+(\I)$ with $\supp(\rho)\subset M_\phi$ and $\rho(\I)=1-\inf_\I \phi$ is a minimizer of the action $\Sact_\I[\,.\,,\phi]$.

Now consider the initial value problem for a given measure $\rho_0\in\M^+(\I)$. Since $\la\L\mu,\mu\ra=(\mu(\I))^2\geq0$ for any signed measure $\mu\in\M(\I)$, the initial data $\rho_0$ is admissible if and only if $\rho_0(\I)\equiv\L\rho_0\leq1$. 
Then for a given external potential $\phi\in\B^+(\I)$ with $\inf_\I\phi\leq1$, the initial value problem has a solution if and only if $\rho_0(\I)\leq1-\inf_\I\phi$ and $\supp(\rho_0)\subset M_\phi$. Namely if $\rho_0=0$, any pair $(\rho,\phi)$ with $\supp(\rho)\subset M_\phi$ and $\rho(\I)=1-\inf_\I\phi$ is a solution. If $\rho_0(\I)>0$, then the measure $\rho=\frac{1-\inf_\I\phi}{\rho_0(\I)}$ is a solution.

Concerning uniqueness, note that in the case $\rho_0(\I)<1-\inf_\I\phi$, we can choose any measure $\mu\in\M^+(M_\phi)$ with $\mu(\I)=1-\inf_\I\phi-\rho_0(\I)$ to obtain the solution $\rho=\rho_0+\mu$ (for example,
we may choose $\mu=(1-\inf_\I\phi-\rho_0(\I))\, \delta_x$ for any $x\in M_\phi$). 
Only in the case $\rho_0(\I)=1-\inf_\I\phi$ there is a unique solution of the initial value problem, namely the trivial solution~$\rho=\rho_0$.

Finally, it is a straightforward observation that the optimization problems (A), (B), (C) and (D) are all solved by choosing any $\phi\in\B^+(\I)$ with $\phi|_{\supp(\rho_0)}=0$ and setting $\rho=\frac{1}{\rho_0(\I)}\rho_0$.

\subsection{The Causal Wedge} \label{sec-causalwedge}
We now analyze a simple system having non-trivial solutions.
Despite its simplicity, this example is instructive because one can compare the different notions of an optimal external potential.
We choose the inner system as three points. Two of these points are space-like separated,
but they are both time-like separated from the third point.
Thus the causal relations coincide with those for three points in Minkowski space
lying at the corners of a wedge with time-like sides.
This is the motivation for the name {\em{causal wedge}}.
More precisely, we let $\I$ be the discrete set $\I=\{1,2,3\}$ and choose the Lagrangian as the matrix 
\beq \L = \bpm 1 & \frac{1}{2} & 0 \\ \frac{1}{2} & 1 & \frac{1}{2} \\ 0 & \frac{1}{2} & 1 \epm\,. \label{ex1-lagr} \eeq
Then any measure $\rho \in \M^+(\I)$ and any external potential $\phi \in \ \B^+(\I)$ can be written as
\begin{align*}
\rho &= (\rho_1, \rho_2, \rho_3) &&\hspace*{-2cm} \text{with} \quad \rho_i\geq0 \\
\phi &= (\phi_1, \phi_2, \phi_3) &&\hspace*{-2cm} \text{with} \quad \phi_i \geq0\,,
\end{align*}
respectively. We choose the initial data~$\rho_0 = (0, \frac{1}{2}, 0)$. Observe that $\L$ is a positive definite operator on $\M(\I)\cong\R^3$ and that $\L\rho_0\leq1$. Therefore, the initial data $\rho_0$ is admissible and its domain of dependence coincides with $\I$ (as follows directly from Proposition~\ref{prop-dependcrit}). Note that the second EL equation \eqref{euler2} holds for any signed measure~$\mu\in\M(\I)$, 
because~$\L$ is positive definite. 
In order to determine the solution of the optimization problems, we now distinguish the cases when the solution of the initial value problem with external potential~$\phi\in\B^+(\I)$ is supported at one, at two, or at all three points, respectively.

\begin{itemize}
\item[(i)] Assume that $(\rho,\phi)\in\mathfrak{S}_\I(\rho_0)$ is such that $\supp(\rho)=\{2\}$. Then $\rho_1=\rho_3=0$, and the EL equation~\eqref{euler1} reads
\begin{align*}
	\frac{1}{2}\rho_2+\phi_1-1 &\geq0 \\
	\rho_2+\phi_2-1 &=0 \\
	\frac{1}{2}\rho_2+\phi_3-1 &\geq0 \,.
\end{align*}
The unique solution of this equation is $\rho=(0,1-\phi_2,0)$, where $\phi_2\in[0,\frac{1}{2}]$ in order to fulfill the constraint $\rho\geq\rho_0$.
Then the action and volume of this solution can be estimated by
\[ \Sact_\I[\rho,\phi]=-(1-\phi_2)^2 \geq -1 \qquad \text{and} \qquad \rho(\I) = 1-\phi_2 \leq 1\,. \]

\item[(ii)] Assume that $(\rho,\phi)\in\mathfrak{S}_\I(\rho_0)$ is such that $\supp(\rho)=\{1,2\}$ (or equivalently, $\supp(\rho)=\{2,3\}$). Then $\rho_3=0$, and the EL equations~\eqref{euler1} read
\begin{align*}
	\rho_1+\frac{1}{2} \:\rho_2+\phi_1-1 &=0 \\
	\rho_2+\frac{1}{2} \:\rho_1+\phi_2-1 &=0 \\
	\frac{1}{2}\:\rho_2+\phi_3-1 &\geq0 \,.
\end{align*}
This system has the unique solution
\begin{align*}
	\rho_1 &= \frac{2}{3}\:(1+\phi_2-2\phi_1) \\
	\rho_2 &= \frac{2}{3}\:(1+\phi_1-2\phi_2)\,, 
\end{align*}
which implies the following estimate for the volume of a minimizer,
\[ \rho(\I) = \frac{2}{3}\: (2-\phi_1-\phi_2) \leq \frac{4}{3} \,. \]
Moreover, the constraint $\rho\geq\rho_0$ imposes relations on $(\phi_1,\phi_2)$,
\begin{align*}
	1+\phi_2-2\phi_1 &\geq 0 \\
	1+\phi_1-2\phi_2 &\geq \frac{3}{4}\,.
\end{align*}
The allowed region for $(\phi_1,\phi_2)$ is a compact convex simplicial subset of $(\R^+)^2$, which is plotted in Figure~\ref{ex1-fig1}.
\begin{figure}[t]
\begin{minipage}[hbt]{6.5cm}
	\centering
	\includegraphics[width=6.5cm]{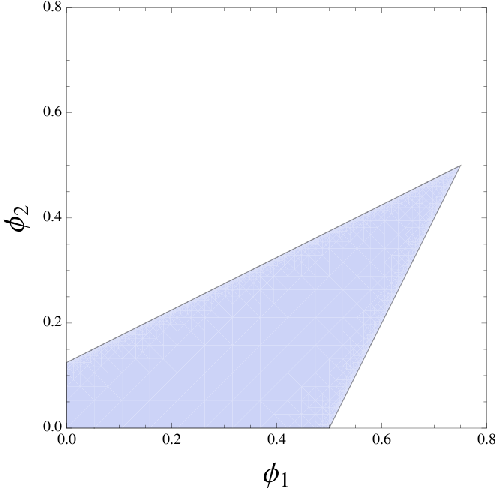}
	\caption{The allowed region for~$\phi$ in case (ii).}
	\label{ex1-fig1}
\end{minipage} 
\hfill
\begin{minipage}[hbt]{6.5cm}
	\centering
	\includegraphics[width=6.5cm]{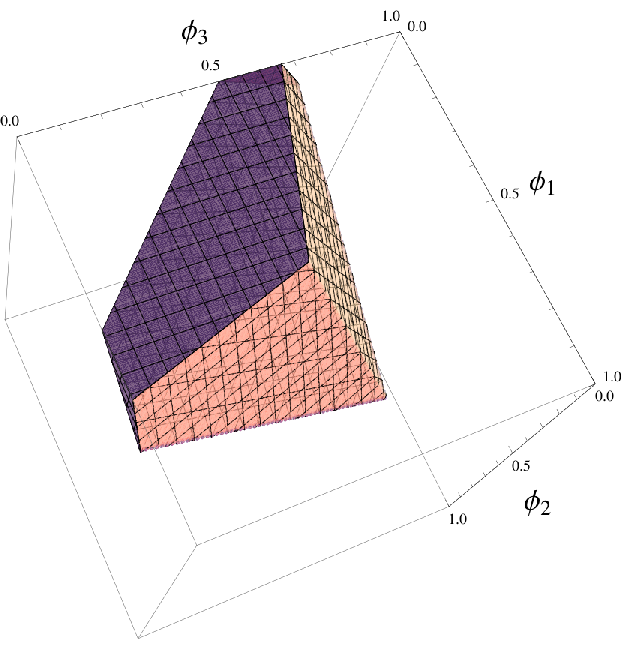}
	\caption{The allowed region for~$\phi$ in case (iii).}
	\label{ex1-fig2}
\end{minipage} 
\end{figure}
The gradient of the action of a minimizer with respect to $\phi$ is given by
$$\nabla_{(\phi_1,\phi_2)}\Sact_\I[\rho,\phi] = \frac{4}{3} \bpm 1+\phi_2-2\phi_1 \\1+\phi_1-2\phi_2 \epm 
\geq \bpm 0\\1 \epm \,, $$
where the last inequality is meant to hold separately for each of the two components.
Thus the minimum of the action in the allowed $\phi$-region lies on the line $\{\phi_2=0\}$, and consequently at $(\phi_1,\phi_2)=(0,0)$. We thus obtain the following estimate for the action of a minimizer,
\[ \Sact_\I[\rho,\phi] = -\frac{4}{3}(1-\phi_1-\phi_2+\phi_1^2+\phi_2^2-\phi_1\phi_2) \geq -\frac{4}{3}\,. \]

\item[(iii)] Assume that $(\rho,\phi)\in\mathfrak{S}_\I(\rho_0)$ is such that $\supp(\rho)=\{1,2,3\}$. Then  the EL equation~\eqref{euler1}reads
\begin{align*}
	\rho_1+\frac{1}{2}\rho_2+\phi_1-1 &=0 \\
	\rho_2+\frac{1}{2}\rho_1+\phi_2-1 &=0 \\
	\rho_3+\frac{1}{2}\rho_2+\phi_3-1 &=0 \,.
\end{align*}
Solving for $\rho$, we obtain
\begin{align*}
	\nonumber \rho_1 &= 1 + \phi_2 - \frac{3}{2}\phi_1 - \frac{1}{2}\phi_3 \\
	 \rho_2 &= \phi_1 + \phi_3 - 2\phi_2 \\
	\nonumber \rho_3 &= 1 + \phi_2 - \frac{3}{2}\phi_3 - \frac{1}{2}\phi_1 \,,
\end{align*}
where $\phi\in\B^+(\I)$ must be chosen such
that the constraints $\rho_1,\rho_3\geq0$ and $\rho_2\geq\frac{1}{2}$ hold.
The allowed region for $\phi$ is a compact convex simplicial subset of $(\R^+)^3$, which is plotted in Figure~\ref{ex1-fig2}.
The constraint $\rho_2\geq\frac{1}{2}$ implies that 
\beq \phi_1+\phi_3\geq\frac{1}{2} \,. \label{constr-1-3} \eeq
Therefore, the volume of a minimizer can be estimated by
\[ \rho(\I) = 2-\phi_1-\phi_3 \leq \frac{3}{2}\,, \]
and the maximal volume is attained along the line $\{\phi_2=0 \:,\: \phi_1+\phi_3=\frac{1}{2}\}$.
The gradient of the action of a minimizer with respect to $\phi$ is given by
\beq \label{grad-iii}
\nabla_{(\phi_1,\phi_2,\phi_3)}\Sact_\I[\rho,\phi] = \bpm 2+2\phi_2-3\phi_1-\phi_3 \\ 2(\phi_1+\phi_3)-4\phi_2 \\ 2+2\phi_2-3\phi_3-\phi_1 \epm \geq \bpm 0\\1\\0 \epm \,,
\eeq
(where the last inequality is again meant componentwise).
Thus the minimum of the action in the allowed $\phi$-region lies in the plane $\{\phi_2=0\}$. 
Setting $\phi_2=0$  and $\phi_1+\phi_3>\frac{1}{2}$, either the first or the third component in \eqref{grad-iii} is strictly positive. Therefore, the constraint~\eqref{constr-1-3} implies that the minimum of the action lies on the line $\{\phi_2=0 \:,\: \phi_1+\phi_3=\frac{1}{2}\}$.
This one-dimensional problem can be solved easily, yielding that the minimal action is attained at the points $\phi=(\frac{1}{2},0,0)$ and $\phi=(0,0,\frac{1}{2})$. Hence the action of a minimizer can be estimated by
\[ \Sact_\I[\rho,\phi] = 2-2(\phi_1+\phi_3)(1+\phi_2)+\phi_1\phi_3+\frac{3}{2}(\phi_1^2+\phi_3^2)+2\phi_2^2 \geq -\frac{11}{8} \,. \]
\end{itemize}
Combining the above cases (i), (ii), and~(iii), we see the following:
The optimization problem~(A) has the two distinct solutions
\beq \rho = \Big(\frac{1}{4},\frac{1}{2},\frac{3}{4} \Big) \;,\; \phi = \Big(\frac{1}{2},0,0 \Big) \qquad \text{and} \qquad
\rho = \Big( \frac{3}{4},\frac{1}{2},\frac{1}{4} \Big) \;,\; \phi = \Big(0,0,\frac{1}{2} \Big) \,. \label{ex1-solA} \eeq
The optimization problem~(C) has a one-parameter family of solutions
\beq \rho_\tau = \Big( \frac{3}{4}-\tau,\frac{1}{2},\frac{1}{4}+\tau \Big) \;,\; \phi_\tau = \Big(\tau,0,\frac{1}{2}-\tau \Big) \qquad\text{with } \tau\in\Big[0, \frac{1}{2} \Big]\,. \label{ex1-solC} \eeq
Hence, neither the problem~(A) nor~(C) has a unique solution on the definite set~$\I$.
Also observe that the two distinct solutions in~\eqref{ex1-solA} both have minimal action as well as maximal volume.

On the other hand, the optimization problem~(D), which maximizes the action in the class of solutions of maximal volume, yields a unique measure $\rho$ according to Theorem~\ref{thm-definiteunique}. This solution is obtained by maximizing the action $\Sact_\I[\rho_\tau,\phi_\tau]$ with~$(\rho_\tau,\phi_\tau)$ given by~\eqref{ex1-solC},
\beq\label{ex1-solD} \rho = \Big( \frac{1}{2}, \frac{1}{2}, \frac{1}{2} \Big) \quad\text{and}\quad \phi = \Big( \frac{1}{4}, 0, \frac{1}{4} \Big)\,.\eeq

In order to see that the solutions of the optimization problem~(B) are also not unique in general, one can add one more point to~$\I$, which is space-like separated from all the other points. Thus $\tilde{\I}=\{\I,4\}$, and the Lagrangian is represented by the matrix
$$\tilde{\L} = \bpm \L & 0 \\ 0 & 1 \epm\,,$$
where $\L$ is the matrix in \eqref{ex1-lagr}. The initial data $\tilde{\rho}_0=(0,\frac{1}{2},0,0)$ is admissible, and its domain of dependence is $\tilde{\I}$. Any solution of the initial value problem in $\tilde{\I}$ is of the form 
\beq \tilde{\rho}=(\rho,1-\phi_4) \quad\text{and}\quad \tilde{\phi}=(\phi,\phi_4)\,, \label{ex2-sol} \eeq 
where $(\rho,\phi)$ is a solution of the initial value problem in $\I$ and $\phi_4\in[0,1]$. 
From the EL equations in the cases~(i), (ii) and~(iii), it follows that~$\max_{\supp(\rho)} \phi\geq\frac{1}{4}$, and hence $\max_{\supp(\tilde{\rho})}\tilde{\phi} \geq \frac{1}{4}$. Thus there is a one-parameter family of solutions of the optimization problem~(B) on the definite set~$\tilde{\I}$, obtained by taking $(\rho,\phi)$ from \eqref{ex1-solD} and choosing~$\phi_4\in[0,\frac{1}{4}]$ in \eqref{ex2-sol}.

\Thanks{{{\em{Acknowledgments:}}
We are grateful to Johannes Kleiner for helpful comments on the manuscript.
A.G.\ would like to thank the German Academic Exchange Service (DAAD) who supported this work by a fellowship within its postdoc program. He is also grateful to the Department of Mathematics at Harvard University for its hospitality while working on the manuscript.}

\providecommand{\bysame}{\leavevmode\hbox to3em{\hrulefill}\thinspace}
\providecommand{\MR}{\relax\ifhmode\unskip\space\fi MR }
\providecommand{\MRhref}[2]{%
  \href{http://www.ams.org/mathscinet-getitem?mr=#1}{#2}
}
\providecommand{\href}[2]{#2}

\end{document}